\documentclass[12pt]{article}

\usepackage{times}
\usepackage{graphicx}
\usepackage{color}
\usepackage{ccaption}
\usepackage{verbatim}
\usepackage{mathtools}
\usepackage{amssymb,amsmath}
\usepackage{multirow}
\usepackage[table,xcdraw]{xcolor}
\usepackage{hhline}
\usepackage{setspace}
\usepackage{versions}
\usepackage{xr}
\usepackage{hyperref}
\usepackage{bm}

\usepackage[labelfont=bf]{caption}

\usepackage [autostyle, english = american]{csquotes}
\MakeOuterQuote{"}
\let\vec\mathbf

\usepackage{setspace}
\newenvironment{sciabstract}{%
\begin{quote} \bf}
{\end{quote}}

\title{Nascent biofilms on soft surfaces}

\author
{Garima Rani$^{1,2}$, G.\ H.\ Philipp Nguyen$^{3}$, Ren\'e Wittmann$^{3,4}$,\\ 
Hartmut L\"owen$^{3^\dagger}$ and Anupam Sengupta$^{1,5^\ast}$
\\
\\
\normalsize{$^{1}$Physics of Living Matter, Department of Physics and Materials Science, }\\
\normalsize{University of Luxembourg, 162 A, Avenue de la Fa\"{i}encerie, L-1511, Luxembourg}\\
\normalsize{$^{2}$ Department of Biochemical Engineering and
Biotechnology,}\\
\normalsize{Indian Institute of Technology Delhi,  Hauz Khas, Delhi 110016, India}\\
\normalsize{$^{3}$ Institut f\"ur Theoretische Physik II: Weiche Materie, Heinrich-Heine-Universit\"at D\"usseldorf,}\\ 
\normalsize{Universit\"atsstra{\ss}e 1, 40225 D\"usseldorf, Germany}\\
\normalsize{$^{4}$ Institute of Safety and Quality of Meat, Max Rubner-Institut, 95326 Kulmbach, Germany}\\
\normalsize{$^{5}$ Institute for Advanced Studies, University of Luxembourg, 2, Avenue de l’Université,}\\
\normalsize{L-4365, Esch-sur-Alzette, Luxembourg}\\
\normalsize{Email: $^\dagger$hartmut.loewen@uni-duesseldorf.de, $^\ast$anupam.sengupta@uni.lu}
}

\date{}
\includeversion{v1}
\excludeversion{v2}

\begin{document} 

%\baselineskip24pt


\maketitle


\begin{sciabstract}
Soft surfaces, spanning vastly different environmental and biomedical settings, are frequently colonised by surface-associated bacteria. Yet, how soft surfaces govern bacterial dynamics and their self-organisation into colonies remains poorly understood. Using experiments and agent-based modelling, we report the self-organisation of bacterial cells into nascent biofilms on soft substrates. By tuning the elastic modulus over two orders of magnitude, we show that the colony morphology, spreading dynamics and collective behaviour depend on the substrate stiffness, wherein softer surfaces promote slowly expanding, geometrically anisotropic, multilayered colonies, while harder substrates drive rapid, isotropic expansion of bacterial monolayers before multilayer structures emerge. Supported a cell mechanical model and two-dimensional agent-based simulations, our results identify that anisotropic drag forces on soft substrates, emerging due to local deformations, underpin colony anisotropy and swift verticalisation. In contrast, reduced drag on hard surfaces allows rapid expansion of monolayers, thus delaying the transition to a multilayer structure. Surface compliance, a key but overlooked determinant of early-stage biofilm development, could be harnessed to engineer biofilm structures and dynamics for nature-inspired and biomedical applications.      

%A major factor underpinning the evolutionary success of bacteria is their ability to form complex surface-associated communities, which extends to a wide range of surfaces which are different in multifarious respects. However, the effect of difference in surface properties on the way cells self-organise in such communities is still not well understood. Here, we study the colony spread and cellular organisation of nonmotile bacteria on substrates with varying agarose concentrations, which consequently have varying agarose concentrations including elastic modulus. We show that colonies show subtle differences in their spreading statistics and colony geometry, when growing on different substrates. This difference is made clear when colony organisation is studied, with colonies growing on softer substrates having markedly different multi-layered geometry, higher boundary instabilities and lesser mono-to-multi layer transition sizes. Our biophysical modelling suggests that drag forces, at different scales, acting on the colony as they spread on the substrate is factor in engendering these differences.
\end{sciabstract}

\maketitle

\section*{Introduction}

From the human gut to soil, and glaciers to radioactive wastes, bacteria are one of the most prolific colonisers, thanks to their ability to successfully inhabit various surfaces \cite{Costerton95,Jin24}. Surface colonisation is crucial for species growth, allowing refuge and protection from detrimental conditions, predation as well as promoting optimal use of resources for maximising growth, cellular behavior and ecology \cite{Katsikogianni20048, Tuson2013, Ahmed2015,Schamberger2023}.

For bacteria, early stages of colonisation is central to their long-term behavior and survival. Surface attachment, followed by the transition from a two dimensional monolayer to a three dimensional multilayer structure \cite{Beloin2008}, often referred to as mono-to-multilayer transition (MTMT) \cite{You2019}, constitute two key steps of bacterial colonisation which have been of keen interest over the last decade, particularly from a mechanobiological viewpoint \cite{You2019,Wingreen2018,Dhar2022,Khan2024}. Depending on the species, surface properties together with local confinements play a crucial role in mediating bacterial colonisation \cite{Hug2017,Courtney2017,Spengler2024,Nezio2024}. A range of surface properties and their effects on bacterial colonisation dynamics have been studied \cite{Tuson2013, Song2015,yin2021}. For instance, surface roughness at micro and nano-scales has been shown to influence bacterial adhesion due to variable contact areas \cite{Mitik2009}. Other such properties include hydrophobicity \cite{Doyle2000, Deepu2023}, surface charge \cite{Hayles2025}, topological features \cite{Sengupta2020,Shimaya2022}, and confinements \cite{Nuno2023}, which influence how bacteria colonise surfaces. Complementing these studies, recent works have indicated the role of surface curvature on bacterial organisation and distribution of biomechanical stresses, particularly during the early biofilm formation. For instance, in both host-microbe settings, as well as in the colonisation of a passive substrate, bacteria-surface interactions can be mediated by the local curvature of the system \cite{Schamberger2023}. Considering that a major source of infections in humans are due to surface transmissions of bacteria, a clear understanding of how various surface properties promote (or inhibit) bacterial colonisation will be key to engineering surfaces to tune bacterial self-organisation into colonies. 

 Compliance, the ability of a surface to deform under stress, has emerged as an important determinant of bacteria-surface interactions, impacting bacterial adhesion, growth, morphological transitions, as well as antibiotic susceptibility \cite{Saha2013,Lichter2008,Song2014,Grant2014, Mu2023}. At high concentration of cells, mechanical constraints imposed by the environment can even lead to glass-like behavior in bacterial populations, as recently observed in dense, quasi-two-dimensional \textit{Escherichia coli} colonies \cite{Lama2024}. Large, colony-scale geometries of bacterial biofilms like those formed by the  \textit{Bacillus subtilis} have been long studied, supported by phase diagrams of their structures with respect to the nutrient medium and substrate stiffness \cite{MATSUSHITA1990}. Recent studies have shown that bacterial behavior encapsulated in soft, 3D hydrogel environments depends on both mechanical and chemical cues \cite{{kandemir2018,Bhusari2022,Lewis2022}}. In mature \textit{E. coli} biofilms, substrate composition and nutrient availability have also been shown to affect internal architecture, including intra-colony channel morphology \cite{Bottura2022}. However, the role of substrate compliance, quantified by the Young's modulus \cite{Rene2024}, specifically during the early stages of a biofilm remains largely unexplored. Majority of existing research focuses on mature colonies comprising thousands of cells, where variables such as nutrient depletion, internal stress regulation, and accumulation of inactive cells introduce considerable complexity, obscuring the direct effects of substrate compliance on nascent biofilms \cite{Allen_2019,Wittmann2023}.

Here, we combine experiments and agent-based simulations to delineate how substrate stiffness impacts the dynamics and cellular organisation within nascent biofilms. Starting with single founder microcolonies of sessile \textit{E. coli} bacteria on nutrient-rich, soft agarose substrates, we study their spatiotemporal dynamics and self-organisation by tuning the "softness" of the substrates, \textit{i.e.}, the elastic modulus, over two orders of magnitude. This is achieved by varying the concentration of low-melting-point (LMP) agarose used in the substrates. We probe the effect of substrate stiffness on the expansion of early-stage biofilms, quantifying the spatiotemporal evolution of the colony geometry, at 2D planar, 3D and fractal dimensions. Our results reveal a subtle influence of substrate compliance in nascent biofilms, which amplify over time to produce marked differences in the colony geometry and mono-to-multilayer transition dynamics over longer timescales. Specifically, our results uncover that stiffer substrates promote morphologically isotropic nascent colonies, with a delayed transition to the multilayered structures; whereas soft, compliant substrates drive anisotropic colony formations, combined with faster 2D to 3D transitions. Agent-based simulations, inspired by our cell mechanical model, suggest that the contrasting observations arise due to higher effective drag forces on soft substrates, due to local deformation that impede the colony expansion as a 2D monolayer. Fractal analysis of the colonies reveals an interesting dichotomy, with colonies on softer substrates displaying higher boundary roughness, driven by mechanical instabilities emerging at the boundary due to the competition of growth-induced expansion and deformation-mediated drag forces. Biophysical modeling highlights the fundamental role of these drag forces in opposing the movement of cells in expanding colonies on soft surfaces, which ultimately determine the resulting variations in colony-scale spreading dynamics. Overall, our results uncover the multi-faceted role of compliant surfaces in mediating the cell-substrate interactions, which could be leveraged to engineer the dynamics and self-organisation of nascent biofilms, and thereby tune their subsequent complex multiscale structures.

%Importance- organisation of cells affected by multiple factors, dependence on substrate key to bacterial colonisation (see review of Tuson-Weitz, Persat on how cells change from planktonic to sessile), difference usually not visible at first glance, necessity of deeper analysis to understand effect. mention effect of adhesion, motivate active-passive. Review of some papers to state results- elife paper of hwa and P.aeruginosa paper, see also Nicolas and other paper for mentioning.

\section*{Results} 
Surface-associated \textit{E. coli} was grown on specially fabricated soft substrates, the stiffness of which was varied via LMP agarose concentrations (Fig.~\ref{fig:FIG1_MAIN}(A) and Materials and Methods), allowing us to tune the Young's modulus, measured using an atomic force microscope (AFM) as reported in Ref. \cite{Rene2024}. While variations in agarose concentrations can result in differences in diffusivity \cite{Bochert2022}, viscosity \cite{agaroseProps} as well as cell surface adhesion \cite{Nezio2024}, we focus here on the impact of substrate compliance on collective dynamics of nascent biofilms.

%As noted, colonies are grown on substrates of varying concentrations of agar, which we will refer to as S1 (0.75$\%$), S2 ($1.5\%$) and S3 (3$\%$). 

\subsection*{Substrate stiffness governs the developmental dynamics of nascent biofilms}

\begin{figure}[ht]
\centering
\includegraphics[width=\columnwidth] {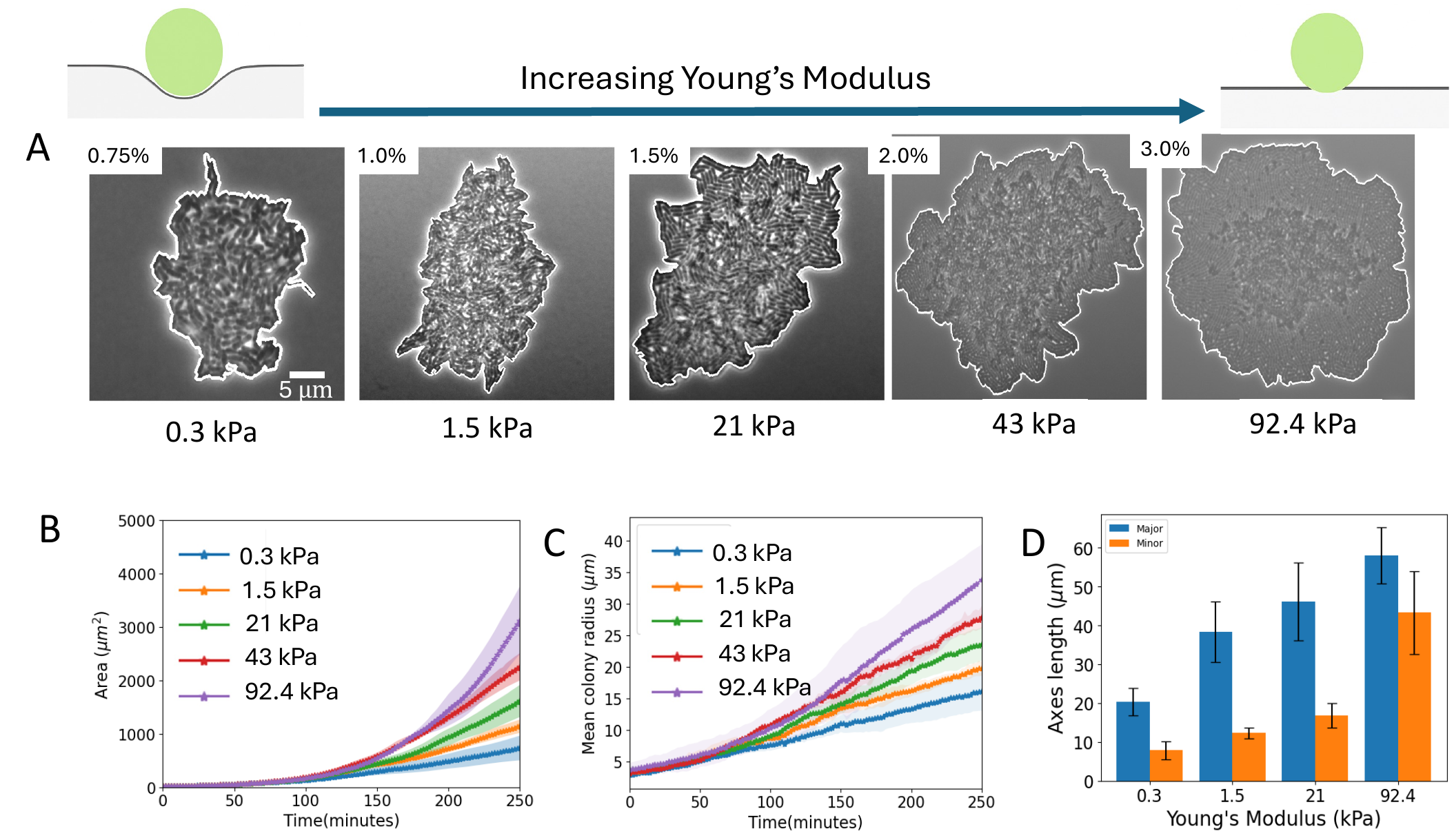}
\caption{\textbf{Morphology of growing \textit{E. coli} colonies on low melting agarose pads, with increasing substrate stiffness.} (A) Snapshots at 6 hours of colony growth on substrates having different concentration of low-melting-point agarose (LMP agarose), as captioned in the images. Here $x\%$ denote $x$ g of LMP agarose in 100 ml of LB media. (B) Colony area is plotted as a function of time. (C) Mean colony radius is plotted as a function of time, (D) The major (blue) and minor (orange) axes of the colonies at MTMT is shown. A significant difference is noted when the Young's modulus of the substrate increases from 21 kPa (low) to 92.4 kPa (high). Data represent at least three biological replicates for each stiffness condition ($n \geq 3$). Statistical difference tested for two sample t-test, p-value $<$ 0.01.}
\label{fig:FIG1_MAIN}
\end{figure}

\begin{figure}
    \centering
    \includegraphics[width=\linewidth]{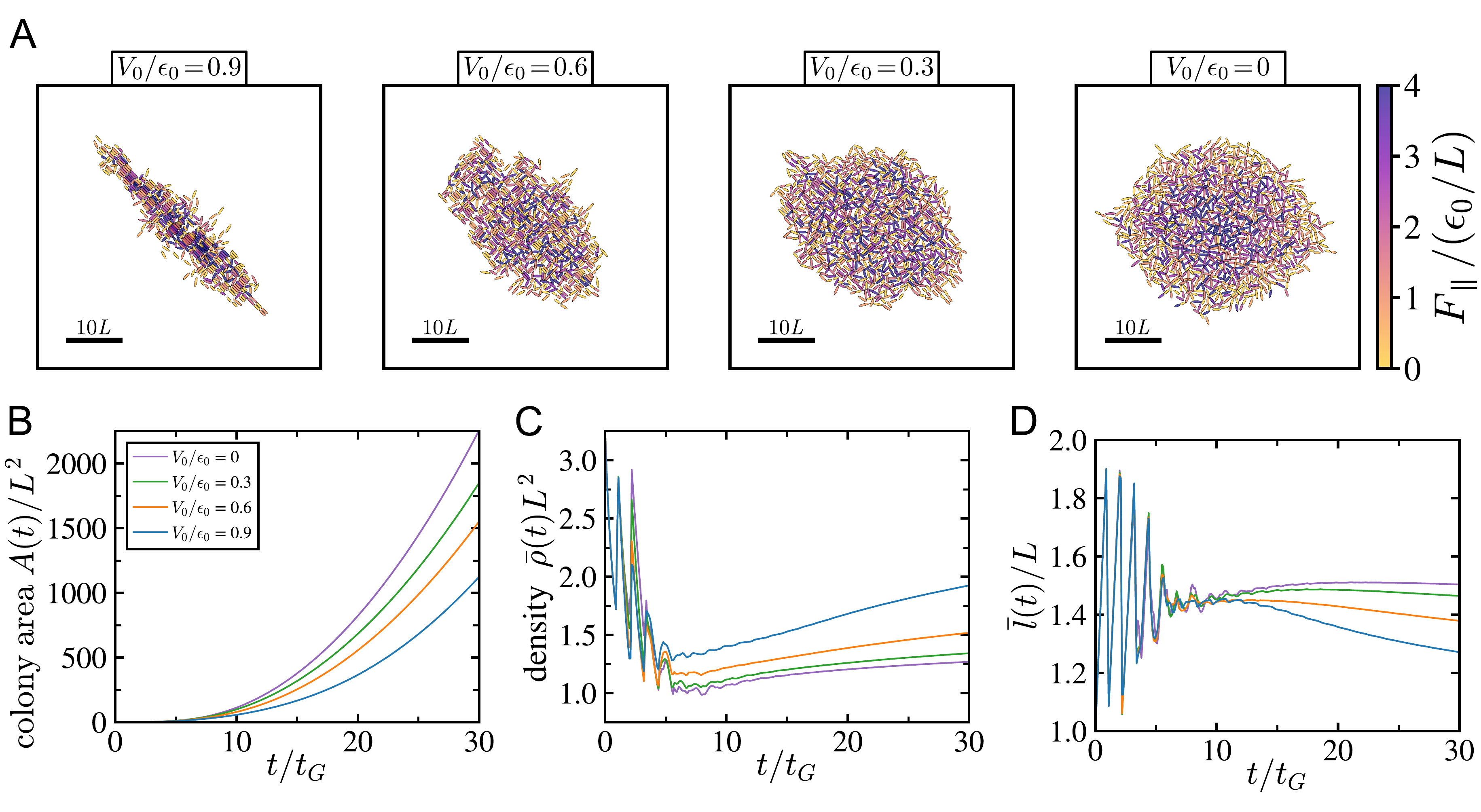}
    \caption{\textbf{Agent-based simulations confirm colony growth dynamics in experiments.} (A) Simulation snapshots at a fixed time point for increasing substrate softness parameter $V_0$, with particles colored according to the magnitude of the parallel force they experience. (B) Colony area $A(t)$ as a function of time for different values of $V_0$. (C) Mean particle density $\bar \rho(t)$ over time, showing increased compaction for larger $V_0$. (D) Mean length $\bar l(t)$ over time, demonstrating that increased density on softer surfaces leads to shorter average cell lengths.
    }
    \label{fig_1}
\end{figure}

Tracking the colony growth in terms of the areal expansion as well as the mean colony radius (Fig.~\ref{fig:FIG1_MAIN}(B) and Fig.~\ref{fig:FIG1_MAIN}(C)) reveals that initially, the colonies show similar growth dynamics, however over longer timescales, the spreading dynamics display marked differences. Colonies expand as a horizontal monolayer on harder substrates, while they develop vertically on softer substrates, as also confirmed by our agent-based simulations of the colony growth dynamics (Fig.~\ref{fig_1}). Consequently, the time for the mono-to-multilayer transition (MTMT) \cite{You2019, Berne2018}, and the doubling time of the projected colony area vary significantly (Fig.~\ref{fig:FIG3_MAIN}). Close to the onset of the MTMT, a significant difference is observed in the primary dimensions of the colonies, i.e., their major and minor axes lengths, as the substrate stiffness is altered. Soft substrates with lower Young's moduli promoted faster MTMT (Fig.~\ref{fig:FIG1_MAIN}(D)), whereas for the colonies growing on harder substrates, the MTMT was delayed. A related trend is depicted by the colony doubling time statistics (Fig.~\ref{fig:FIG3_MAIN}(A)), confirming that over time, colonies on harder substrates spread faster horizontally, while those on compliant surfaces have a slower expansion. 

The experimental results are supported by our two-dimensional agent-based model, where bacterial cells are treated as elongating, dividing, and interacting rod-shaped particles, subject to forces from their neighbours (Fig.~\ref{fig_1}). A key feature of the model is an effective attraction, $V_0$, a measure for the substrate softness, that mediates the interaction between the cells \textit{in silico} depending on the deformable substrates (see Materials and Methods for a detailed description). Molecular dynamics (MD) simulations conducted over a range of $V_0$ (larger $V_0$ value corresponds to a soft surface and vice versa) reveal that the collective behavior and morphology of the growing colonies depend on the strength of this attraction. Figure~\ref{fig_1}(A), representating simulation snapshots for increasing hardness of the substrates (i.e., decreasing values of $V_0$), captures a clear trend: as the substrate-mediated attraction decreases, the resulting colonies become larger and more isotropic in their geometry. Furthermore, the rate of increase of the projected area is significantly suppressed for larger $V_0$ (Fig.~\ref{fig_1}(B)), in agreement with the trend in the mean length (Figs.~\ref{fig_1} and \ref{fig:FIG6_MAIN}), and total cell number $N(t)$ (cf. Supplementary Fig. 1(A)), thus confirming that substrate-mediated attraction limits the colony's ability to proliferate and expand, in excellent agreement with the experimental results.

%Further details on the simulation parameters can be found in the Supplementary Information.

%The overall growth dynamics were characterized by tracking the colony area and the number of constituent particles over time.

\subsection*{Substrate stiffness impacts cellular organization in nascent biofilms}

\begin{figure}
\centering
{\includegraphics[width=\columnwidth]{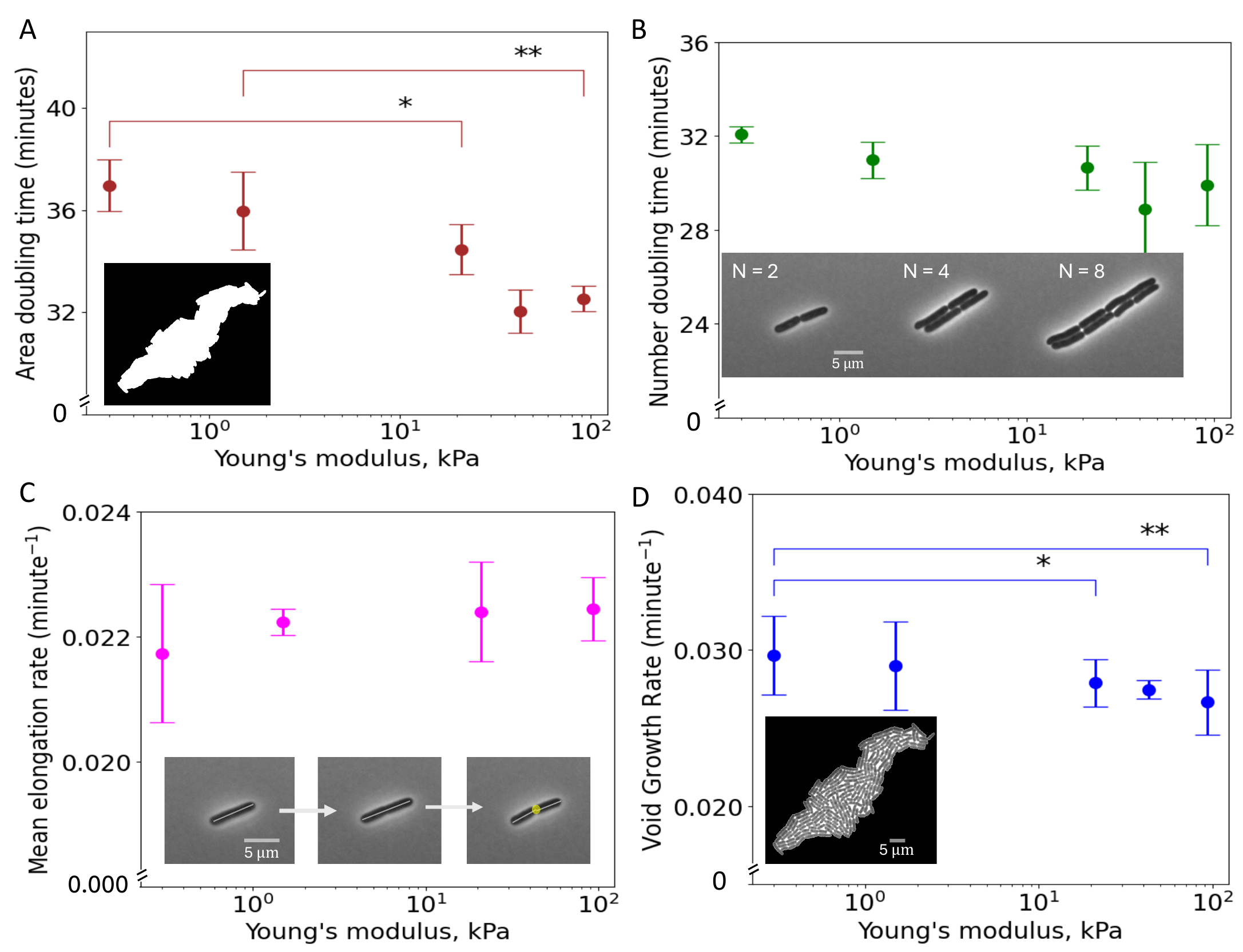}}
\caption{\textbf{Substrate stiffness impacts the growth of bacterial colonies.} Various doubling times are reported here: (A) area doubling time, (B) cell number doubling time, (C) Mean elongation rate of cells, and (D) Growth rate of voids in the colony. Comparisons are made to check the variation in values when stiffness of the substrate is changed. A significant difference in area doubling time and void growth rate values is observed when the Young's modulus varies from from low to high values. Data represent at least three biological replicates per stiffness condition for statistical analysis. Asterisks  correspond to a specific level of significance: two sample t-test, p-value$<$ 0.05 ($^*$); and p-value $<$ 0.01 ($^{**}$).}
\label{fig:FIG3_MAIN}
\end{figure}

 To analyse the mechanistic role of substrate compliance that drives the contrasting observations reported above, we first plot the population doubling time. Interestingly, this shows little variation across the different substrates (Fig.~\ref{fig:FIG3_MAIN}, as also indicated by the statistics of the doubling time and cellular elongation rates (Fig.~\ref{fig:FIG3_MAIN}(B, C)), which likewise show no significant variation across the different cases. Thus, at the level of single cells, we do not observe a statistically significant difference between cell-level growth and division rates due to variation of the substrate compliance, in agreement with previous studies \cite{Gomez23}. Thus, the variations at the colony-scale reported in previous Section, are likely driven by the organizational and geometric variations within the expanding bacterial colonies. To probe this, we next focus on the impact of substrate stiffness on the cellular organization in colonies and the overall colony geometry.

\subsubsection*{Bulk morphology of growing colonies depend on the substrate stiffness}

Cell arrangement in bacterial colonies depend on local parameters including temperature, for instance, spatial confinements and fluid flow \cite{Dhar2022, Nuno2023, Rani2024, Jingyan2021, Hartmann2019}. However, if, and to what extent, compliance of the underlying substrates impact the growth dynamics as well as collective self-organisation remain unexplored \cite{You2018, Dell2018}.  At the colony-scale, the major and minor axes of the colonies show similar dynamics over short timescales, however they diverge over longer times, depending on different stiffness conditions (Supplementary Fig. 2). While the major and minor axes increase comparably for harder substrates, this tendency stalls for soft substrates where the major axis outweighs the minor axis. Furthermore, colonies growing on substrates with $2\%$ and $3\%$ agarose concentrations show nearly similar growth trends, although the stiffness varies by more than 100\%. The minor axis of the colonies shows similar growth statistics for the softer substrates, however as the stifness is increased, it approaches closer to the major axis, nearly matching up in the case of hardest substrate tested here, i.e., with the highest agarose concentration ($\sim 3\%$ concentration). Taken together, colonies growing on hard substrates show isotropic morphology (Fig.~\ref{fig:FIG1_MAIN}), as also confirmed by the colony aspect ratios, i.e., the ratio of the two axes (Supplementary Figs. 2, 3 and 4(A)).

\begin{figure}[ht]
    \centering
    \includegraphics[width=\linewidth]{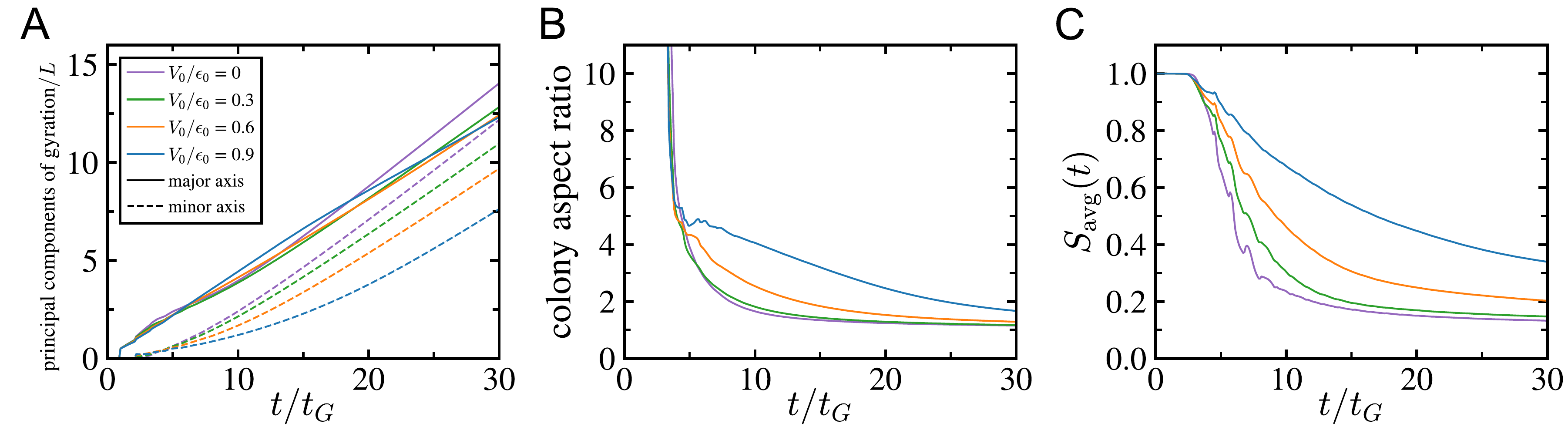}
    \caption{
    \textbf{Emergence of colony anisotropy captured by agent-based simulations.} (A) The two eigenvalues of the gyration tensor of colonies over time. The minor eigenvalue $\lambda_{\text{min}}$ is strongly suppressed for larger $V_0$, i.e., softer substrates. (B) Colony aspect ratio, calculated from the eigenvalues of the gyration tensor, as a function of time. (C) Average local nematic order parameter $S_\text{avg}(t)$ over time, showing that orientational alignment is better preserved on softer surfaces.
    }
    \label{fig_2}
\end{figure}

Our simulated colonies closely support the experimental data. We characterize the emergent anisotropic shape of the simulated colonies, by extracting the weighted gyration tensor
\begin{equation}
    S_{\alpha\beta} = \frac{\sum_{i=1}^{N(t)}l_i(r_{i,\alpha} - r_{c,\alpha})(r_{i,\beta} - r_{c,\beta})}{\sum_{i=1}^{N(t)}l_i} \,,
\end{equation}
where $r_{i,\alpha}$ is the $\alpha$-component of the position of particle $i$ with length $l_i$ and $r_{c,\alpha}$ is the $\alpha$-component of the colony's center of mass.
The eigenvalues of this tensor $\lambda_{\text{maj}}^2$ and $\lambda_{\text{min}}^2$, correspond to the squared principal semi-axes of the colony.
As shown in Fig.~\ref{fig_2}(A) and the Supplementary Fig. 3 (for the experimental data), the major axis $\lambda_{\text{maj}}$ grows comparably for all values of $V_0$, however, the minor axis $\lambda_{\text{min}}$ is strongly suppressed on softer surfaces.
The resulting attractive forces primarily act to confine the colony along its shorter dimension, preventing it from expanding into a circular shape.

This differential increase of the principal axes results in a highly anisotropic colony morphology.
We quantified this using the colony aspect ratio $\lambda_{\text{maj}}/\lambda_{\text{min}}$ in Fig.~\ref{fig_2}(B).
After an initial transient phase, the aspect ratio for all colonies decreases from an initially end-to-end arrangement of cells.
For larger $V_0$, however, this decrease is much slower, leading to a significantly more elongated colony at later times.
This result precisely reproduces the experimental observation of more anisotropic colonies on softer substrates (See Supplementary
Fig. 4A).
It is important to note that this quantity provides a meaningful measure of colony shape only after a sufficient number of particles ($N \gtrsim 10$) have formed.
In the earliest stages of growth, when particles are arranged in a nearly perfect line, the minor principal axis $\lambda_{\text{min}}$ is close to zero, causing the aspect ratio to diverge theoretically.
To complement this physical measure, we also calculate the aspect ratio of the colony's minimal bounding box, a purely geometric quantity that does not suffer from this divergence (Supplementary Fig. 1(C)).

The underlying mechanism for this persistent anisotropy is the orientational alignment of the particles.
We measured the local nematic order parameter, which quantifies the degree of alignment of a particle with its immediate neighbors.
Figure~\ref{fig_2}(C) shows that for larger $V_0$, the average local nematic order $S_\text{avg}(t)$ remains significantly higher throughout the simulation.
The substrate-mediated potential wells provide an energy incentive for neighbouring cells to align, stabilizing the nematic order against both thermal fluctuations and the significant mechanical perturbations caused by cell division events.
Since cells elongate and divide along their major axis, this robust local alignment directly translates into anisotropic colony growth, as the colony preferentially expands along the direction of mean alignment.
This robust orientational alignment, ultimately, provides the foundation for the emergence of a globally anisotropic colony.

Compliance of the underlying substrates impact the local cell-level arrangements. As cells in growing colonies are constantly pushed around due to growth-induced activity \cite{You2018, Dell2018}, spatiotemporal arrangement can be characterised in terms of microdomain size distributions, the local angular and polar distributions \cite{You2018, Shimaya2022}, as well as the formation of genealogical enclaves and their intermixing characteristics \cite{Rani2024}. By quantifying the formation of local voids, \textit{i.e.}, cell-free regions, within the colonies due to growth-induced reorganisation (Fig.~\ref{fig:FIG3_MAIN}(D)), we observe a clear trend: higher void growth rate in case of soft substrates, while harder substrates promote lower void growth rate. On softer substrates, voids are larger as well as they grow faster, suggesting that the packing density of the cells in a colony can be tuned by the substrate stiffness, and could be leveraged as a metric for predicting relative stiffness of different substrates.

To understand the internal structure, we analyzed the mean particle density.
As shown in Fig.~\ref{fig_1}(C), the mean number density $\bar \rho(t) \coloneqq N(t)/A(t)$ initially decreases as the first few cells grow and divide into free space.
However, it then increases as the colony becomes crowded.
This effect is far more pronounced for larger $V_0$, where the stronger attractions pull the particles closer together, leading to a much denser colony.
A direct consequence of this high-density packing is observed in the mean particle length (Fig.~\ref{fig_1}(D)).
While particles on harder surfaces (small $V_0$) reach an average length of approximately $1.5L$, particles in the confined environment of a softer surface (large $V_0$) are subject to a larger mechano-response of its neighbours \cite{Wittmann2023}, resulting in a steady decrease in the average particle length over time.
That is the reason why the colony grows slower on the softer surfaces.
Note that the initial fluctuations of the mean number density $\bar \rho$ and the mean length $\bar l$ are a consequence of synchronous cell cycles during the early generations.

\begin{figure}
\centering
\includegraphics[width=\columnwidth] {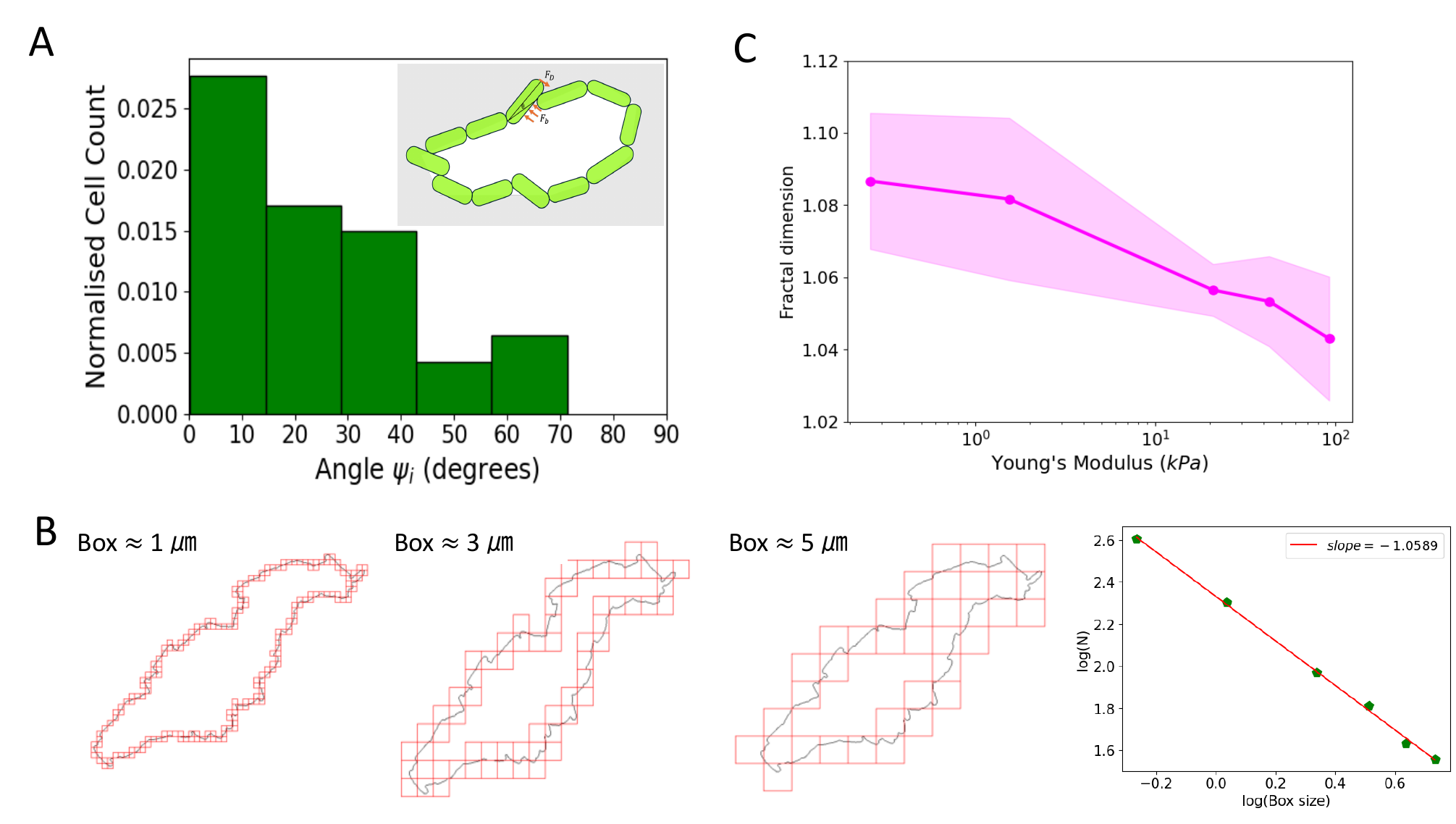}
\caption{\textbf{Boundary roughness of the growing colonies quantified by fractal dimension.} (A) Histogram of the acute angle between the boundary tangent and bacteria orientations. Most cells tend to align with the tangent of the colony boundary. However tangential order breaks at the colony boundary due to imbalance between growth-driven expansion forces (due to the expanding colony) and the opposing drag force from the substrate, leading to orientational disorder at the boundary. High fractal at softer substrate (Inset). (B) Box counting algorithm for computing the fractal dimension of the colony boundary: the colony boundary coordinates are computed and the image is then divided into grids, with grid size progressively increased, and the number of boxes intersecting the colony boundary counted. The fractal dimension in each case is computed by finding the slope of logarithm of the number of boxes containing part of the boundary and logarithm of the box size. (C) The fractal dimension is plotted (mean and standard deviation), for colonies at around transition time for increasing Young's modulus of the LMP agarose substrate.}
\label{fig:FIG4_MAIN}
\end{figure}

\subsubsection*{Fractal dimension of boundary roughness scales inversely with substrate stiffness}

The interactions between cells and substrates are arguably maximised at the boundary of the colony, with the colony actively pressing its way forward to expand. The moving bacterial front on a soft surface may be impeded by substrates, with the boundary functioning as an "active-passive interface" \cite{Zhang2022}. This gives rise to striking interfacial behavior, including a proliferation of topological defects closer to the outer periphery \cite{Sengupta2020, Rani2024, Amin2016}. Consequently, we hypothesise that the substrate compliance will impact the cell arrangement at the colony boundary, even more than it does at the center of the colony. To test this hypothesis, we quantify the local orientational order on the colony boundary. Cells on the boundary are largely tangential to the boundary (Fig.~\ref{fig:FIG4_MAIN}(A)), as has been noted in interfacial regions in multiple contexts \cite{You2018, Amin2016, Franco2017}. However, disorder occurs when some cells are pushed out of the tangential configuration and adopt orientations angular to the cell boundary, resulting in a characteristic "rough" colony boundary. We perform fractal analysis of the boundary (see Fig.~\ref{fig:FIG4_MAIN}(B), and Materials and Methods), allowing us to go beyond the Euclidean geometric methods typically used for 
morphological analysis. Fractal analysis, on the other hand, has been increasingly used in multiple reports to analyze and quantify morphological complexity of cells and their aggregations \cite{Porter91, fractalbook}. In our case, the disorder in the orientation of cells at the colony boundary results in an oscillating boundary curve, whose "magnitude" of oscillation and consequently, the degree of orientational disorder, can be captured by the fractal dimension of the boundary curve. Our data shows that fractal dimension of the boundary curves of colonies decreases with increasing substrate stiffness (Fig.~\ref{fig:FIG4_MAIN}(C)). The results indicate that the higher fractal dimension, capturing a higher degree of disorder at the boundary, is observed for colonies on soft substrate. Fractal dimension closer to 1, as seen in the case of hard substrates, suggest a smoother curve with most of the cells aligned tangentially with the colony boundary. This is also indicated by the distribution of the angular difference (between the colony boundary tangent and cell orientation at MTMT, Supplementary Fig. 5), which peaks at very low angles for colonies growing on stiffer substrate. 

Thus, we conclude that a higher degree of orientational disorder occurs at the boundary of colonies grown on softer substrates, in contrast to the higher bulk nematic order measured within these colonies. The irregular boundary morphology likely offers a stronger impediment to the colony expansion, as the local deformations result in a higher effective drag. The resulting stresses lead to a closer packing of cells, associated with higher orientational alignment. In the following, we investigate how these effects facilitate the out-of-plane extrusion of cells, and thereby tune the transition dynamics of cellular monolayers to multilayers.

%\section*{Simulation Results}
%\subsection*{Cell-based simulations}

\subsection*{Dynamics of mono-to-multilayer transition on compliant substrates}

The mono-to-multilayer transition (MTMT), a phenomena that is widely conserved across bacterial species and morphotypes, occurs when the active in plane stresses within the colony become high enough to overcome restoring forces, driving cells to climb on top of the base layer \cite{Wingreen2018,You2019}, leading to the formation of layered 3D structures \cite{Jingyan2016, Hwa2019, Yunker2024}. Time-lapse imaging of the growing colonies indicates striking differences in the 3D structural geometry of colonies growing on soft versus hard substrates (Fig.~\ref{fig:FIG1_MAIN}(A)). For the softest substrates considered here, the second layer spontaneously adopts the shape and size as that of the underlying monolayer, making them optically indistinguishable. In contrast, colonies growing on harder substrates, have a distinct second layer, growing over the monolayer. Furthermore, the layers are relatively circular in shape (aspect ratio $\approx$ 1), and the 3D geometry vis-a-vis the initial planar geometry, is clearly distinguisable, owing to the optical contrast between the layers. For colonies growing on low agarose substrates, i.e., on soft substrates, the second layer is spread out almost completely over the base layer, while for high agarose case, layers growing after colony transition into 3D structure have a much smaller spread compared to the base layer. In general, for all cases, colonies tend to adopt morphologies which become circular with time, particularly after the transitioning to multi-layered structures (Supplementary Fig. 4(A)). This shows that the compliance of the underlying substrate modulate 3D morphology, from the addition of layers after the planar monolayer to the overall aspect ratio. While previous studies have looked at large-scale 3D structures of bacterial colonies growing on substrate properties \cite{MATSUSHITA1990, Geisel2022, Jingyan2023}, the structural complexities have remained overlooked. Notably, for large 3D colonies, similar observations have been made, wherein harder substrates have been reported to promote colonies which are flatter and better packed, while on softer substrates, the colonies adopt loose, dome-like geometries \cite{Jingyan2023}, in agreement with our observations here, and as also suggested by our void analysis (Fig.\ref{fig:FIG3_MAIN}(D)). Interestingly, analogous behaviour has been observed in wetting on substrates and in cellular aggregates undergoing wetting transitions, with droplets on soft substrates displaying more compact, partial wetting morphologies instead of full spreading while rigid substrates promote complete wetting \cite{wetting1, wetting2}.

\begin{figure}
\centering
\includegraphics[width=\columnwidth] {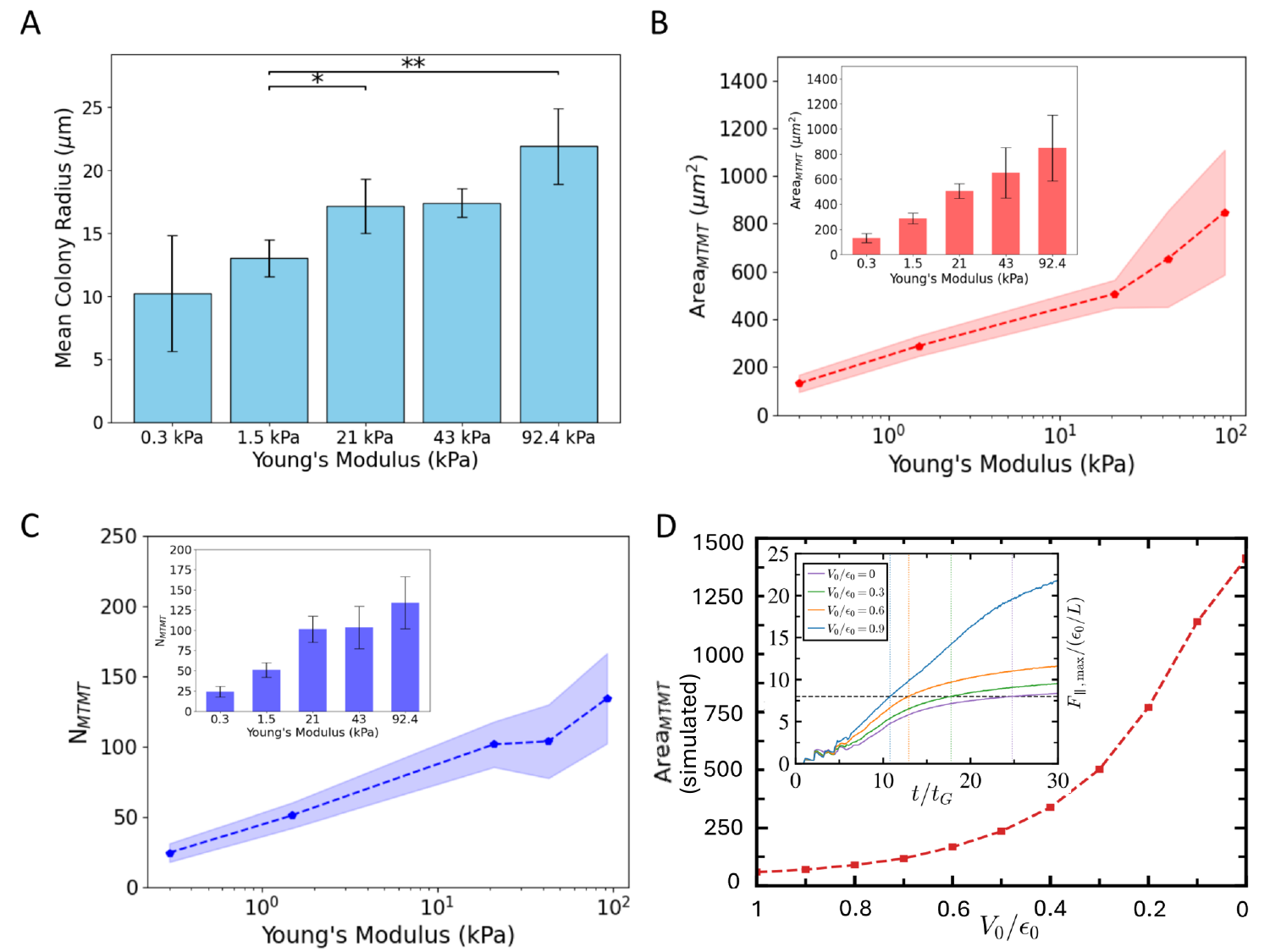}
\caption{\textbf{Growth dynamics and geometric parameters as the colony transitions to the multilayered morphology.} Variation of the (A) mean colony radius, (B) colony area, (C) cell number count, and (D) Dependence of the total area of the colony at the mono-to-multilayer transition, as a function of the substrate compliance. Inset shows the Maximum parallel force in the colony as a function of time
for different values of $V_0$. The horizontal dashed line indicates the
critical buckling threshold,while the vertical dotted lines mark the
transition points.}
\label{fig:FIG5_MAIN}
\end{figure}

Colonies growing on soft substrates exhibit an early onset of the mono-to-multilayer transition, as compared to those growing on harder substrates (Fig.~\ref{fig:FIG1_MAIN}). To understand this, we probe multiple spatiotemporal characteristics related to the 3D transition across various substrates. First, we compare the size of the colony at the onset of the MTMT. We observe that colonies growing on softer substrates attain MTMT at smaller colony sizes (measured in terms of colony radius, colony area, as well as the number of cells), relative to substrates with higher agarose concentrations (Fig.~\ref{fig:FIG5_MAIN}(A-C)). This is also reflected in the orientational order parameter $S_\text{avg}(t)$ over the colony area, revealing a more heterogeneous distribution MTMT (Supplementary Figs. 6 and 7, see also Ref. \cite{Rani2024}). Numerically, the colony size at the transition increases with increasing substrate stiffness. The difference is comparable also when the colony area and the cell numbers are compared. For instance, colonies on the softest substrates are around 4-5 times smaller at MTMT as compared to the colonies growing on the hardest substrate considered here. It is pertinent to note here that, as observed above, the colonies in different cases show similar areal growth statistics, initially (Fig.~\ref{fig:FIG1_MAIN}(B,C)), which in fact is related to their structural transition to a 3D structure. Indeed, colonies growing on soft substrates spread at a similar rate as those growing on harder substrates till they transition to a multilayer structure. Further, as noted, the colonies on all the substrates get more circular with time, though circularity of the colony around the MTMT also shows a dependence on substrate properties, showing an increasing trend with the substrate stiffness (Supplementary Fig. 4(B)).

While our simulation model is strictly two-dimensional, it allows us to estimate the point in time at which a real colony would begin to transition to a three-dimensional, multilayered structure. This mono-to-multilayer-transition, determined by an interplay of growth-mediated extensile forces and the surface interactions, occurs when the compressive forces on a cell reaches a critical threshold \cite{You2019,Rene2024}. We thus track the parallel force experienced by any cell in the colony (see Fig.~\ref{fig_1}(A)), and compare the critical threshold across different substrate conditions (see inset, Fig.~\ref{fig:FIG5_MAIN}(D), and Materials and Methods section for the simulation details).
Our results clearly confirm the experimental trends that on softer surfaces (large $V_0$), stronger attractions lead to higher internal stresses, causing the maximum parallel force to reach the critical threshold much earlier in the colony's development. We quantify this by plotting the transition time, particle number, and colony area at the point when this threshold is crossed (see Fig.~\ref{fig:FIG5_MAIN}(D) and Supplementary Fig. 8). For stiffer substrates (smaller $V_0$), the onset of the multilayer transition is significantly delayed, occurring at later times when the colony is composed of more particles and covers a larger area. We may as well recall that the boundary of the colonies growing on softer substrates appear more disordered, suggesting that cells get locally perturbed out of their positions more easily on softer substrates. Taken together, this presents an apparent dichotomy, which we clarify below by integrating information at both the colony and the individual scales.

\begin{figure}
\centering
\includegraphics[width=\columnwidth] {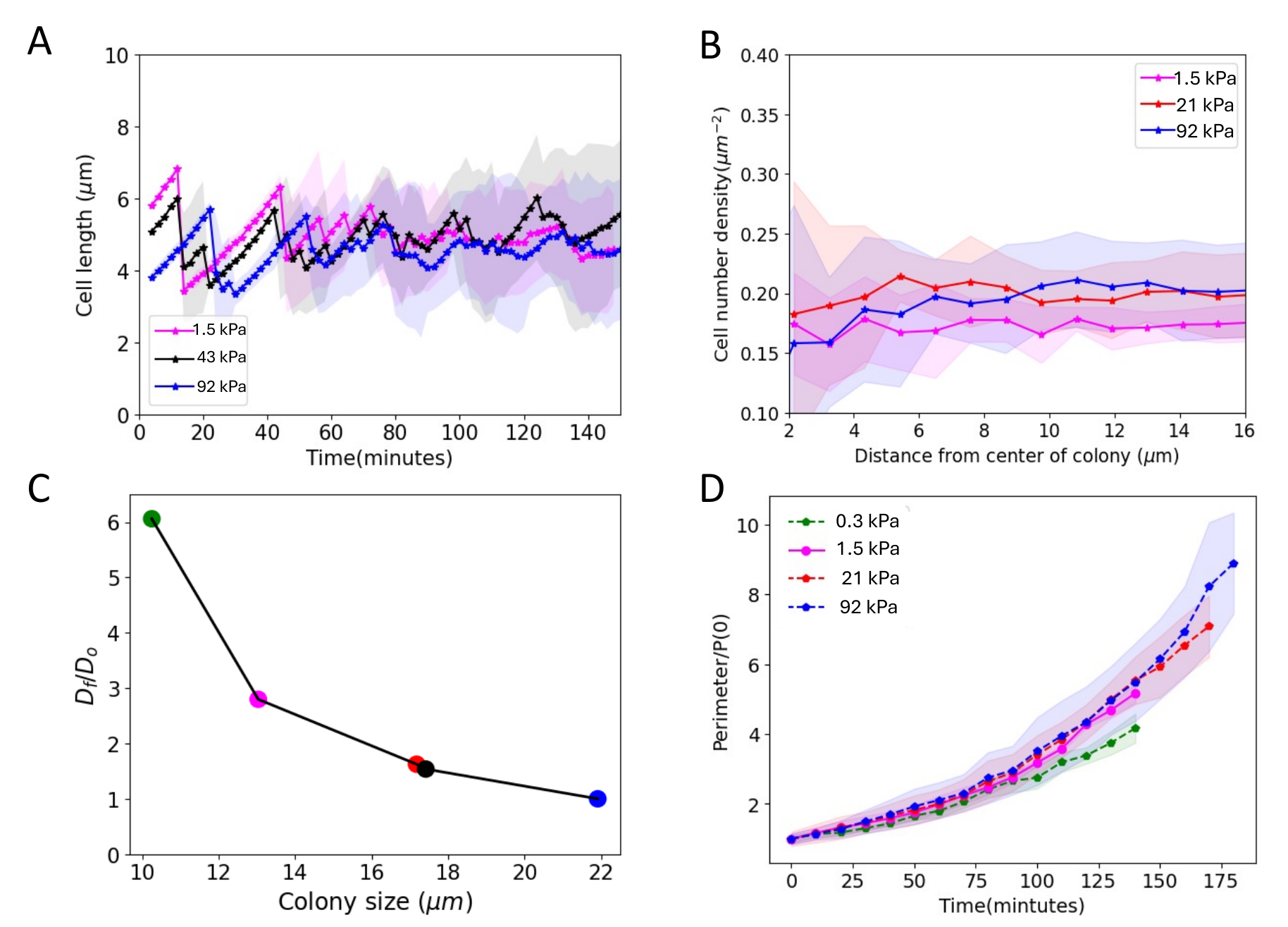}
\caption{\textbf{Spatiotemporal changes in the colony due to the variation of the substrate stiffness.} (A) Mean cell length of cells in the colonies is plotted as a function of time, (B) Spatial variation in the cell number density is plotted as a function of distance from the colony center, (C) Estimated drag parameter $D_f$ (normalized by the drag parameter for the case of colonies growing on  hardest substrate, having 3$\%$ LMP agarose concentration) is plotted as a function of colony size at the transition to multilayered growth. Here, each point correspond to different Young's modulus values, and (D) Normalized colony perimeter is plotted with time for different substrates with varying Young's modulus.}
\label{fig:FIG6_MAIN}
\end{figure}

%In the case of substrate with the lowest agar concentration, we observe an interesting feature of cells show a pattern of a going under neighbouring cells at very small colony sizes. 

\subsection*{Substrate-mediated drag forces impede colony expansion on soft substrates}

We develop a simple biophysical model of cells growing on a soft substrates to shed light on possible mechanisms through which cell-substrate interactions modulate cellular organization and large-scale structures within the colonies. Our model allows to qualitatively understand the biomechanical underpinnings of the observed phenomena, including the dichotomy in degree of orientational disorder at the colony edge as well as the rapid nucleation of multi-layered colonies on soft surfaces. We hypothesise two primary modes of cell-surface interactions in a developing bacterial colony: emergent drag forces due to local substrate deformations, which impede the displacement of cells on the substrate; and the cell-surface adhesion, which remain relatively constant between \textit{E. coli} and LMP agarose, as reported previously \cite{Rene2024}. For the timescale of interest here (up to MTMT), the variation of adhesive forces between cells and the agarose surface is negligible \cite{Rani2024, Rene2024}, thus we focus on the emergent drag forces on the expanding colonies.

%\begin{figure}
%\centering
%\includegraphics[width=4in] {Figures/Stiffness_update-fig/fig6.pdf}
%\caption{\textbf{Emergence of effective drag forces on growing colonies.} Relative sinking depth estimated for colonies growing on the various substrates, with dots indicating the Young's modulus values. (Inset) Illustrative sinking of colony into substrate, with depth estimated using Hertzian contact.}
%\label{fig:FIG7}
%\end{figure}

The cell size distribution together with the cell-surface adhesion determine the out-of-plane extrusion \cite{You2019}. As shown in our recent work, the adhesive component for \textit{E. coli} does not vary with the Young's modulus \cite{Rene2024}, and so is the cell size distribution (Fig~\ref{fig:FIG6_MAIN}A). So, here we estimate how the variation in the substrate compliance results in the difference of drag forces experienced by the growing colonies. For simplicity, we take the average colony size at MTMT as the representative colony size, since the internal stress at the center exceeds the critical threshold.

Considering a 1D chain model for bacteria, with length $L$ that extends from $-L/2$ to $L/2$, we quantify the relative differences in colony size at MTMT, and derive the relative difference in the drag force that is experienced by the colonies spreading over various substrates. The stress evolution within the 1D colony follows as: 

\begin{equation}
       \partial_x \sigma_x + D_fv_x=0\,,
\end{equation}

\noindent where $\sigma_x$ denotes the growth-induced normal stress along the $x$ axis. The colony density evolves as

\begin{equation} \label{eq_rhoc}
      \partial_t\rho_c+\partial_x(\rho_c v_x) =k_c\rho_c\,,
\end{equation}

\noindent where $\rho_c$ is the density, $v_x$ denotes the direction of velocity along the $x$, and $k_c$ is the growth rate. The drag parameter, $D_f$, quantifies the scale at which colony expansion is impeded due to the soft substrate. With $\rho_c$ being constant spatiotemporally throughout the colony (Fig.~\ref{fig:FIG6_MAIN}(B)), the first term in Eq.~\eqref{eq_rhoc} vanishes, and under the free boundary conditions $\sigma(-L/2)=\sigma(L/2)=0$, we get

\begin{equation}\sigma_x (x)= \frac{k_cD_fL^2}{8}\bigg(1-\bigg(\frac{2x}{L}\bigg)^2\bigg)\,,\end{equation}

\noindent where $L$ denotes the size of the colony. Stress is maximum at $x=0$ and so, for the critical stress value $\sigma_c$, we can derive colony size at transition as

\begin{equation}\label{eq:Dragf}
    L_c=\sqrt{\frac{8\sigma_C}{k_cD_f}}\,.
\end{equation}

\noindent As noted above, colonies spread at a similar rate for all cases initially, so the drag parameter $D_f$ for different cases can be compared by relative differences in the colony sizes at MTMT, using Eq. \ref{eq:Dragf} (Fig.~\ref{fig:FIG6_MAIN}(C)). We see that drag is much higher in case of colonies growing on softer substrates, indicating a stronger impediment to the colony expansion, thus ultimately resulting in a more rapid transition from monolayer to multilayer structures. 

We extend the drag force analysis for the growth of a 2D colony, using Rayleigh's drag equation

\begin{equation}\label{eq:drag force}
    F_D=\frac{1}{2} \rho C_d v^2 A\,,
\end{equation}
where $F_D$ is the drag force experienced by the moving object, $\rho$ is substrate density, $C_d$ is the drag coefficient which depends on the speed, $v$, and cross-sectional contact area, $A$, of the moving object, and the substrate properties. Equation \ref{eq:drag force} suggests that the cross-sectional area plays a key role, since it determines the magnitude of the dynamic resistance encountered. A combination of the colony perimeter and the local deformation of the substrate contributes to the overall drag. In contrast to a 1D colony which expands only in two directions (front and rear of the chain), such that its contact area is always of the same magnitude as for individual cells; the 2D colony expansion front increases constantly, leading to a rapid increase in the contact area with time, as the perimeter increases with time (Fig.~\ref{fig:FIG6_MAIN}(D)). To arrive at a relative magnitude of the contact-induced deformation, we consider the 2D Hertzian contact model, under a spherical shape assumption (radius, $R$), to estimate the contact-induced sinking depth (Supplementary Fig. 9): 

\begin{equation}
    \delta \approx (\frac{9F^2}{16E_{cs}^2R})^{\frac{1}{3}}\,,
\end{equation}

\noindent where $F$ is the normal force due to the cell mass, and $E_{cs}$ is the effective cell-substrate elastic modulus \cite{Landau86}. The effect of substrate stiffness is encoded in the effective cell-substrate elastic modulus $E_{cs}$ which is given as 

\begin{equation}
    E_{cs}=\frac{4E_cE_s}{3(E_c+E_s)}\,,
\end{equation}
where $E_c$ and $E_s$ denote the elastic modulus of cell and substrate respectively, with the Poisson ratio of 0.5 for the cell and the substrate \cite{Rene2024}. However, as $E_c$ values, given by the elastic modulus of bacterial cells is around $\sim 20-50$ MPa \cite{Yao1, Tuson2012}, the $E_c>> E_s$, giving $E_{cs}\approx 4E_s/3$. 

Thus, estimating the relative sinking depth for similar sized colonies on the various substrates, we observe that the sinking depth varies inversely with substrate stiffness (Supplementary Fig. 9). Now, the contact area for a single cell of length $l$ is then given as $\sim l\delta$ (for a typical cell, this will be $\sim 0.002-0.03~\mu$m$^2$), while for a colony with an expanding front of perimeter $P$, the contact area will be $\sim P\delta$ (for a colony of 100 cells of radius 15 $\mu m$, this will be $\sim 0.5-3~\mu$m$^2$). As noted, this encapsulates the difference in the 1D vis-a-vis the 2D case, with the exponential growth in the 2D case creating a key difference in the way drag forces act on growing bacterial colonies (Fig.~\ref{fig:FIG6_MAIN}(C)). Taking the results together, the difference between resulting individual and collective drag forces emerge as the main driver of the evolution of colony morphology. 

In general, cells at the colony boundary are aligned tangentially, and occasionally break this symmetry and adopt skewed orientations at the colony-substrate interface. The skewed orientations are triggered due to local instabilities at the boundary. Say, a tangentially aligned cell is rotated out of position by the force at the colony-substrate border, reorienting it to an angle $\theta$ relative to the colony boundary Fig.~\ref{fig:FIG4_MAIN}(C). The local force due to the expanding bulk, denoted as $f_b$, acts to move cells out of tangential orientation while the opposing force due to drag, acts against the motion of the cell as it rotates to change its orientation. Likewise, the torque $\tau_b=lf_b\cos\theta$ is opposed by the drag torque, which is given by $\tau_d=lF_D$. Thus, cells will rotate out of the tangential alignment whenever the local bulk forces exceed $F_D$, which is directly proportional to $\rho A$, where $\rho$ is the substrate density and $A$ is the cross-sectional contact area (see Eq. \ref{eq:drag force}). However, in this scenario, the cross-section area enters at the single cell level, and is thus bounded by $l\delta$, which is is much less compared to the cross-section area, $P\delta$, of the colony. So, we can conclude that orientational disorder at the boundary is decided at the individual scale. As bulk forces dominate for softer substrates which have lower density, allowing cells to move out of tangential position with greater ease in these cases, the orientation of cells is more disordered at the colony boundary. On the other hand, as the colony grows in 2D, the cross-section area grows exponentially, so combined with the greater depth $\delta$ of cells sinking into the softer substrates, this results in higher drag force and thus, larger forces in bulk promoting a faster transition into multi-layered colony morphologies.

\section*{Summary and discussion}

Growth of bacterial colonies on surfaces is one of the most important factors underpinning the resilience and ubiquity of bacteria in a wide variety of ecological settings. Still, a clear understanding of the way bacterial cells interact with soft surfaces while growing into a colony, has remained lacking. By varying the stiffness of substrates, achieved by tuning concentrations of low melting agarose, we uncovered multiple organizational features, from individuals to colony-scales, which were impacted by the substrate stiffness. While the colony expansion was similar over the short time-scales, colonies growing on harder substrates expanded faster and spread over a wider horizontal area. While the growth and division dynamics of cells remain statistically comparable across different substrates, the self-organization of cells within these colonies is impacted. Softer substrates promoted higher growth of voids in colonies as compared to those growing on harder substrates. Further, analyzing the transition from planar to multi-layer 3D morphologies, confirmed that the colonies growing on softer substrates undergo a more rapid transition to the multi-layer structures, i.e., they also expand vertically rather than purely horizontally. Upon transition, the 3D colonies showed marked differences, with colonies growing on harder substrates exhibiting a clear demarcation between the isotropic circular layers, while those on softer substrates displayed heterogeneous structures, with the second layer closely following the spread of the base layer. Biophysical modeling revealed the key role of effective drag forces experienced by the colonies as they spread on different substrates. Drag forces acting at both the individual-as well as the colony-scales, promote faster transition to multilayer on softer substrates, in agreement with previous reports \cite{Rana2017}. Our experimental observations were confirmed by computer simulations of a cell-based model, which incorporates these drag forces via an effective attraction. 

Our work delineates mechanistically the wide-ranging impact of substrate compliance on the dynamics and self-organisation of bacteria across scales, which may be of particular relevance to the distribution and composition of microbiomes associated with humans. Studies elsewhere have also demonstrated that bacteria are capable of causing considerable deformations on soft host surfaces, allowing to develop a mechanical model of bacterial infection \cite{Persat2020}. While bacterial colonies can change their material properties under mechanical stresses, such changes depend also on substrate properties like agar concentration \cite{Kochanowski2024}, highlighting that mechanical cross-talks between hosts and microbial agglomerations offer diverse fundamental as well as applied prospects \cite{Sengupta2020}. The current study, focusing on early stages of colony formation initiated by a single founder cell, offer important distinctions with previous reports which have generally looked into the large-scale cell agglomerations, typically comprising thousands of cells. For instance, growth has been observed to slow down as the substrate stiffness increased \cite{Saha2013, Song2014, Little2019, Yan2017}, though in some other cases it has been observed to increase with stiffness \cite{Asp2022}. In fact, studies in fungal systems have shown that stiffer substrates can promote faster surface expansion, suggesting that mechanical constraints can directly modulate growth rates \cite{Yang2024}. For large colonies, nutrient availability as well as secondary metabolites can play an important role in determining the colony behavior \cite{Allen_2019}, however at the scales of we have studied here, nutrient availability is clearly uniform across the colony \cite{Dhar2022}. Further mechanistic insights could be obtained by accounting for the agarose microstructures, which might additionally mediate the local deformation as a function of the concentration, and thereby play a role in the colony expansion dynamics \cite{agaroseProps}. Finally, for motile bacterial species, additional biophysical considerations may be relevant, for instance, the cell-surface adhesion properties, as in the case of \textit{Chromatium okenii} from our past work \cite{Rene2024}, can be very different from the non-motile \textit{E.coli}. This opens up an important avenue for future work, particularly in the context of plankton-to-biofilm transitions on soft surfaces. 

\newpage

\section*{Methods}

\subsection*{Cell-based simulations}

To complement our experimental observations and investigate the underlying physical mechanisms driving colony morphogenesis on soft substrates, we developed a two-dimensional agent-based model.
In this model, individual bacteria are represented as elongated particles whose dynamics are governed by overdamped Langevin equations.
The model incorporates cell growth, division, and physical interactions, including particle-particle repulsion and a substrate-mediated attraction that depends on the effective softness of the surface.
It is important to note that this two-dimensional model is designed to describe the initial expansion of the colony as a monolayer, prior to any potential buckling or transition to multilayered, three-dimensional growth.

Each bacterium $i$ in the system is described by its state vector, which includes its center of mass position $\vec{r}_i = (x_i, y_i)$, its orientation angle $\phi_i$, and its length $l_i$.
While the length $l_i$ is a dynamic variable, the particle width $W$ is fixed at all times and identical for all cells.
The dynamics of these variables are governed by the following set of overdamped Langevin equations
\begin{align}
\frac{\mathrm{d}\vec{r}_i}{\mathrm{d}t} &= -\mu_t \, \nabla_{\vec{r}_i} U + \sqrt{2D_t}\bm{\xi}_t(t) \;,\\
\frac{\mathrm{d}\phi_i}{\mathrm{d}t} &= - \mu_r \, \frac{\partial U}{\partial \phi_i} + \sqrt{2D_r}\xi_r(t) \;,\\
\frac{\mathrm{d}l_i}{\mathrm{d}t} &= G - \mu_l \frac{\partial U}{\partial l_i} + \sqrt{2D_l}\xi_l(t) \;.
\end{align}

Cell growth is explicitly modeled with the (average) elongation rate $G > 0$ and cell division, which drives the system out of equilibrium.
A cell $i$ divides when its length $l_i$ reaches the division length $2L$.
Upon division, the mother cell is removed and replaced by two new daughter cells.
The daughter cells inherit the orientation $\phi_i$ of the mother cell, are initialized with a length of $L$, and are placed symmetrically around the mother cell's final position along its major axis shifted by $L/2$.
A single cell with initial length $L$ divides after the bare generation time $t_G \coloneqq L/G$.
Hence, at time $t$ there are $N(t)$ particles, labeled $i=1, \ldots, N(t)$, and this number increases with time.

The terms $\bm{\xi}_t(t)$, $\xi_r(t)$, and $\xi_l(t)$ represent uncorrelated Gaussian white noise of unit variance, modeling thermal fluctuations.
Their magnitudes are determined by the corresponding diffusion coefficients $D_t$, $D_r$, and $D_l$.

The total potential energy of the system $U$ generates the interaction forces and torques that mediate the particles' dynamics.
The term $-\nabla_{\vec{r}_i} U$ is the force that displaces the particle's position, while $-\partial U / \partial \phi_i$ is the torque that changes its orientation, where $\mu_t$ and $\mu_r$ are the translational and rotational mobilities, respectively.
Crucially, our model also includes a mechanical response of the cell's growth to its environment.
The term $-\partial U / \partial l_i$ represents the force acting on the particle's length, which resists elongation when the particle is under compression from its neighbors.
This mechanical feedback directly counteracts the elongation rate $G$, thus coupling the cell's effective growth rate to the local mechanical stress.
The strength of this so-called mechano-response is quantified by $\mu_l$, which is related to the substrate friction \cite{Wittmann2023}.

The total potential $U = U_{\text{int}} + U_{\text{ext}}$ is the sum of two contributions: a particle-particle interaction potential $U_{\text{int}}$ and an effective external potential mediated by the substrate $U_{\text{ext}}$.

\begin{enumerate}
    \item \textbf{Particle-particle interaction ($U_{\text{int}}$):}
    To account for short-range repulsion between cells, we employed an anisotropic Gaussian repulsive pair potential \cite{berne1972gaussian}.
    The total interaction potential is the sum over all unique pairs of particles $U_{\text{int}} = \sum_{i<j} U_{ij}^{(\text{rep})}$.
    The specific form of the potential $U_{ij}$ is given by
    \begin{equation}
        U_{ij}^{(\text{rep})}(\vec{r}_{ij},\phi_i,\phi_j,l_i,l_j) = 4\epsilon_0 \exp\left[ -\left( \frac{2|\vec{r}_{ij}|}{\sigma(\tilde \phi_i,\tilde \phi_j,l_i,l_j)} \right)^{\!2} \right]
    \end{equation}
    with the potential strength $\epsilon_0$ and the anisotropy function \cite{persson2011simple, persson2012note}
    \begin{equation}
        \sigma(\tilde \phi_i,\tilde \phi_j,l_i,l_j) = W \left\{ 1 + \frac{1}{2}\left[ \left(\frac{l_i}{W} - 1\right)\cos(\tilde \phi_i) + \left(\frac{l_j}{W} - 1\right)\cos(\tilde \phi_j) \right] \right\} \;,
    \end{equation}
    where $\tilde \phi_{i,j}$ are the angles with respect to the distance vector $\vec{r}_{ij} = \vec{r}_{i}-\vec{r}_{j}$ between particles $i$ and $j$.

    \item \textbf{Substrate-mediated interaction ($U_{\text{ext}}$):}
    The crucial element of our model is the representation of the soft substrate.
    We assume that each bacterium deforms the soft surface, creating a potential well that reflects its own anisotropic shape.
    This creates an effective, many-body attraction.
    We modeled this by having each particle $j$ generate an inverted anisotropic Gaussian potential field.
    The potential experienced by a particle $i$ due to particle $j$ is given by
    \begin{equation}
        U_{ij}^{(\text{att})}(\vec{r}_i) = -V_0 \exp\left[ -\left( \frac{(x'_{ij})^2}{2\sigma_{l,j}^2} + \frac{(y'_{ij})^2}{2\sigma_{W}^2} \right) \right] \,,
    \end{equation}
    where the total external potential on particle $i$ is $U_{\text{ext}}(\vec{r}_i) = \sum_{j \neq i} U_{ij}^{(\text{att})}(\vec{r}_i)$.
    The coordinates $(x'_{ij}, y'_{ij})$ are the coordinates of particle $i$ in the reference frame of particle $j$.
    This frame is centered on particle $j$ and rotated by its orientation angle $\phi_j$, with the transformation from the lab frame given by
    \begin{equation}
        \begin{pmatrix} x'_{ij} \\ y'_{ij} \end{pmatrix} = \begin{pmatrix} \cos\phi_j & \sin\phi_j \\ -\sin\phi_j & \cos\phi_j \end{pmatrix} \begin{pmatrix} x_i - x_j \\ y_i - y_j \end{pmatrix} \;.
    \end{equation}
    The standard deviations of the potential well $\sigma_{l,j}$ and $\sigma_{W}$ are directly proportional to the dimensions of the particle $j$ creating the well; specifically, $\sigma_{l,j}$ scales with the particle's dynamic length $l_j$ and $\sigma_W$ with its constant width $W$.
    The parameter \mbox{$V_0 \geq 0$} is the amplitude of the interaction, defining the depth of the potential well.
    In our framework, $V_0$ serves as a proxy for the softness of the substrate.
    A hard, non-deformable surface is modeled by a small $V_0$, while a softer, more deformable surface corresponds to a large $V_0$.
    Note that because this potential acts on the center of mass of particle $i$, it does not exert any direct torques on its orientation ($\partial U_{\text{ext}} / \partial \phi_i = 0$) or forces on its length ($\partial U_{\text{ext}} / \partial l_i = 0$).
\end{enumerate}

\subsubsection*{Threshold stress for out-of-plane transition}
The mono-to-multilayer-transition begins when the lifting torque due to cell-cell repulsion is larger than the restoring torque due to cell-substrate adhesion.
When the compression forces between cells exceed the critical value
\begin{equation}
    F_\text{crit} = \frac{F_\text{adh}}{1 + l/W} \;,
\end{equation}
that point in time is the transition time $t_\text{MTMT}$ \cite{You2019}.
With an adhesion force of $F_\text{adh} \approx 0.3\,$nN \cite{Rene2024}, $\epsilon_0 \approx 10^4 k_\text{B}T$ and $l/W \approx L/W = 2.5$, the estimated critical force is $F_\text{crit}/(\epsilon_0/L) \approx 8$.
Since only compression forces along the long axis of a cell would lead to buckling into the third dimension, we track the parallel force $F_\parallel$ experienced by any cell in the colony (see Fig.~\ref{fig_1}(A)) and compare the maximal value to the critical force threshold $F_\text{crit}$ (see inset, Fig.~\ref{fig:FIG5_MAIN}(D)).

\subsection*{Cell culture and experimental protocol}

We used non-motile strain of Gram negative bacteria {\it{E.coli}}, namely NCM3722 delta-motA. Single colonies were initially streaked on LB-agar plates and incubated at $30 ^\circ$ C. Individual colonies were picked using a sterile inoculation loop and transferred to liquid LB medium in a shaking incubator set at 170 rpm. Cultures were grown to late exponential phase, as monitored by optical density (OD), and then diluted 1:1000 into fresh LB medium. These cultures were grown for an additional 1.5 to 2 hours to reach a mid-exponential growth phase.  

To prepare the LB-agarose gel substrate, low-melting-point agarose (LMP, Preparative Grade for Large Fragments ($>$ 1,000 bp), Promega, USA) was mixed with LB medium to create an agarose-LB solution. The solution was poured into a Gene-Frame (spacer) adhered to a clean microscope slide. After pouring, the gel was allowed to cool and solidify. Approximately 1 to 1.5 $\mu$ L of bacterial culture was inoculated on the substrate and single cell-to-colony dynamics was observed using time-lapse microscopy. 

The Young's modulus of the LMP agarose gel substrates were measured using atomic force microscopy (AFM), using contact mode and fitted with standard elasticity models, as detailed in our previous work \cite{Rene2024}.

\subsection*{Image acquisition}
The time-lapse imaging protocol, with $2$ minute interval, was designed to capture bacterial growth and colony formation dynamics. LMP agarose concentrations was varied from $0.75\%$ to $3\%$, in the prepared substrates. Multiple regions on the substrate were imaged for every experiment. At least three biological replicates were performed for each substrate stiffness, as well as multiple technical replicates were captured for each biological replicate, covering different positions on the sample. The growing colonies were observed using phase contrast microscopy (Olympus IX83, $60\times$/$100 \times$ oil objective, and imaged using Hamamatsu ORCA-Flash camera), within a temperature-controlled incubator. The experiments were performed at $30 ^\circ$ C to match the growth conditions of the bacterial strain. Single bacterial cells acted as monoclonal nucleation sites, expanding horizontally on the nutrient-rich LB-agarose layer. Initially, the colony expanded horizontally, as a monolayer in two dimensions. Over longer timescales, the bacterial monolayer transformed into multiple layers, thereby developing into a three dimensional structure over multiple generations. Further details on the experimental protocol can be found in our previous studies \cite{Dhar2022, Rani2024}.

\subsection*{Image analysis}
Phase-contrast images were processed using Fiji:ImageJ \cite{Schindelin2012}, Ilastik \cite{Berg2019}, and custom python scripts developed in-house. The image data were pre-processed, involving brightness adjustment and background noise subtraction (via ImageJ), followed by additional noise reduction and filtering using python-OpenCV. A subset of images was used to train Ilastik’s pixel classifier to separate cells from the background, and the classifier was applied to the remaining frames via batch processing, iterating for improved segmentation. Segmentation was performed until the colony transitioned to the third dimension, tracked carefully by monitoring changes in the image contrast within the bulk of the colony. Cell length was calculated as the distance between the centroid and the poles (extreme ends), validated using OpenCV ellipse fitting for the major axis. Colony boundaries were detected using OpenCV’s \textit{findContours} and \textit{flood-filling} routines, allowing us to obtain the boundary coordinates, colony area as well as the perimeter. Void growth rate was determined by establishing the time series of the area of voids (regions within the colony boundary which were not occupied by cells) in the colony and measuring their growth rates. Statistics reported in our analyses and figures were calculated across all replicates.

\subsection*{Computations}

\subsubsection*{Colony-scale geometric features}
Colony geometry and shape parameters were calculated by fitting a rotating bounding rectangle to the colony boundary and computing its major and minor dimensions, as described in Supplementary Figs. 2-4. Colony shape was assessed by calculating aspect ratio ($R_{major}/R_{minor}$) and circularity (a measure of compactness, defined as the ratio of the area of the colony to the area of a circle with the same perimeter). Furthermore, for each set of experiments across substrates with different Young's moduli, we extract the distance of the colony boundary from the corresponding center. We present this as the probability distribution of distance from the center of the colony over $360^{\circ}$ angle, as shown in Supplementary Fig. 10.

\subsubsection*{Fractal analysis}
The fractal dimension of the colony boundary was calculated using the box-counting method. First, the grayscale image was converted into a binary image, and the colony boundary was extracted. The box-counting algorithm was applied by dividing the image into grids of varying sizes, and counting the number of boxes which intersected with the boundary. This process was repeated with progressively smaller grid sizes (down to approximately $0.5 ~\mu m$). A plot was generated of the logarithm of grid size (box size) versus the logarithm of the number of boxes which intersect the boundary. The slope of the plot corresponds to the fractal dimension of the boundary (Fig.~\ref{fig:FIG5_MAIN}B).

\subsubsection*{Cell orientation relative to the colony boundary}
Segmented binary images of bacterial colonies were analyzed to study local cell orientation near the colony boundary. Cell centroids within a defined distance from the edge of the colony were extracted. A closed B-spline curve was fitted through these centroids to represent the colony boundary locally. For each cell, the acute angle between its major axis and the tangent to the B-spline at the nearest point was computed. These angles measured the alignment between cell orientation and the local boundary direction. The probability density (see Supplementary Fig. 5) of the angle between the tangent at the colony boundary (green arrows) and the cell orientation (red arrows, with black dots representing the centroids) reveals that most cells tend to align tangentially along the colony boundary. The substrate stiffness however influences the extent to which this local alignment is achieved: cells growing on stiffer substrates exhibit a higher degree of tangential alignment, whereas low alignment is observed in case of cells growing on softer substrates, resulting in higher boundary roughness.

\subsubsection*{Nematic order parameter}
Cell orientation in phase-contrast images was computed using the structure tensor method, with local averaging performed using a Gaussian window approximately one-quarter the size of a single cell. Orientation angles were extracted via eigenvalue analysis of the tensor. The local nematic order parameter, $S_\text{avg}(t)$, was calculated within square regions R (containing ~3–4 cells) using a moving grid approach. Only pixels within bacterial cells were considered. The nematic order parameter for colonies growing on substrates with different Young's moduli is shown in Supplementary Figs. 6 and 7. Details of the algorithm is provided in Ref. \cite{Rani2024}.

\subsection*{Statistical tests}
Statistical analysis was performed using two-sample t-tests across different substrate stiffness values, given by their Young's modulus. Data are presented as mean ± standard deviation (SD). Statistical significance was defined as $p < 0.05$. The asterisks on the plots indicate significance levels: $p < 0.05$ ($^*$) and $p < 0.01$ ($^{**}$). All analyses were conducted using Python (SciPy and NumPy).

\newpage

\section*{Acknowledgments}
 
 We gratefully acknowledge the support from Human Frontier Science Program Cross-Disciplinary Fellowship (LT 00230/2021-C to G.R.) and the Institute for Advanced Studies, University of Luxembourg (AUDACITY Grant: IAS-20/CAMEOS to A.S.). We thank René Riedel for the fruitful discussions and support with the Young's modulus measurements. H.L. and R.W. acknowledge funding from the German Research Foundation (DFG) within the SPP 2265 (projects HL 418/25-2 and WI 5527/1-2). A.S. thanks the Luxembourg National Research Fund for the ATTRACT Investigator Grant (A17/MS/11572821/MBRACE) for supporting this work.

\bibliography{paper.bib}

\bibliographystyle{unsrt}

\end{document}


\baselineskip24pt


\maketitle

\vspace{10pt}

%\section*{Acknowledgments}
%Include acknowledgments of funding, any patents pending, where raw data for the paper are deposited, etc.
\begin{center}

\section*{Supplementary Information}

\end{center}

The supplementary file contains 9 Supplementary videos covering experiments and simulations, additional details on the methods employed for the experiments as well as simulations reported here, and 10 Supplementary figures.

\newpage

\subsection*{Supplementary videos}

\textbf{Raw data of colonies growing on substrates of varying stiffness}

\begin{enumerate}
    \item \textit{1\_RAW\_0.3kPa.avi}: Raw data of a colony growing on a substrate with stiffness \textbf{0.3 kPa}.
    \item \textit{2\_RAW\_1.5kPa.avi}: Raw data of a colony growing on a substrate with stiffness \textbf{1.5 kPa}.
    \item \textit{3\_RAW\_21kPa.avi}: Raw data of a colony growing on a substrate with stiffness \textbf{21 kPa}.
    \item \textit{4\_RAW\_43kPa.avi}: Raw data of a colony growing on a substrate with stiffness \textbf{43 kPa}.
    \item \textit{5\_RAW\_92.4kPa.avi}: Raw data of a colony growing on a substrate with stiffness \textbf{92.4 kPa}.
    
\end{enumerate}

\textbf{Simulated colonies growing on substrates of varying stiffness}

\begin{enumerate}

     \item \textit{6\_SIM\_00.mp4}: Simulated colony growth on the hardest surface considered \textit{in silico}.

      \item \textit{7\_SIM\_03.mp4}: Simulated colony growth on surface with lower hardness considered \textit{in silico}.

       \item \textit{8\_SIM\_06.mp4}: Simulated colony growth on the soft surface considered \textit{in silico}.

        \item \textit{9\_SIM\_09.mp4}: Simulated colony growth on the softest surface considered \textit{in silico}.
    
\end{enumerate}

%\begin{figure}
%\centering
%\includegraphics[width=3in] {Figures/Stiffness_update-fig/suppl_Rene.pdf}
%\caption{\textbf{Young's modulus of substrate.} Young’s modulus of the LM agarose substrates in water as a function of agarose concentration.}
%\label{fig:FIG1_YM}
%\end{figure}

\newpage

\subsection*{Simulation details}
The overdamped Langevin equations of motion were integrated using a forward Euler-Maruyama method with a fixed time step of $\Delta t = 10^{-4}t_G$. Each simulation was initialized with a single particle of length $L$ with a random orientation and position. A total of 4096 independent simulation runs were performed for each parameter set to ensure robust statistics.
The following parameters were used for the simulations presented in the main text:
\begin{itemize}
    \item particle width $W = 0.4L$
    \item substrate-mediated potential shape $\sigma_{l,j}^2 = l_j^2/2$ and {$\sigma_{W}^2 = W^2/2$}
    \item translational mobility $\mu_t = \frac{L^2}{\epsilon_0 t_G}$
    \item rotational mobility $\mu_r = \frac{2}{\epsilon_0 t_G}$
    \item mechano-response strength $\mu_l = \frac{L^2}{2\epsilon_0 t_G}$
    \item translational diffusion $D_t = 10^{-4} \frac{L^2}{t_G}$
    \item rotational diffusion $D_r = 2 \times 10^{-4} \frac{1}{t_G}$
    \item magnitude of the fluctuations of the elongation rate $D_l = 10^{-4} \frac{L^2}{t_G}$
\end{itemize}

\begin{figure}[ht]
    \centering
    \includegraphics[width=0.9\linewidth]{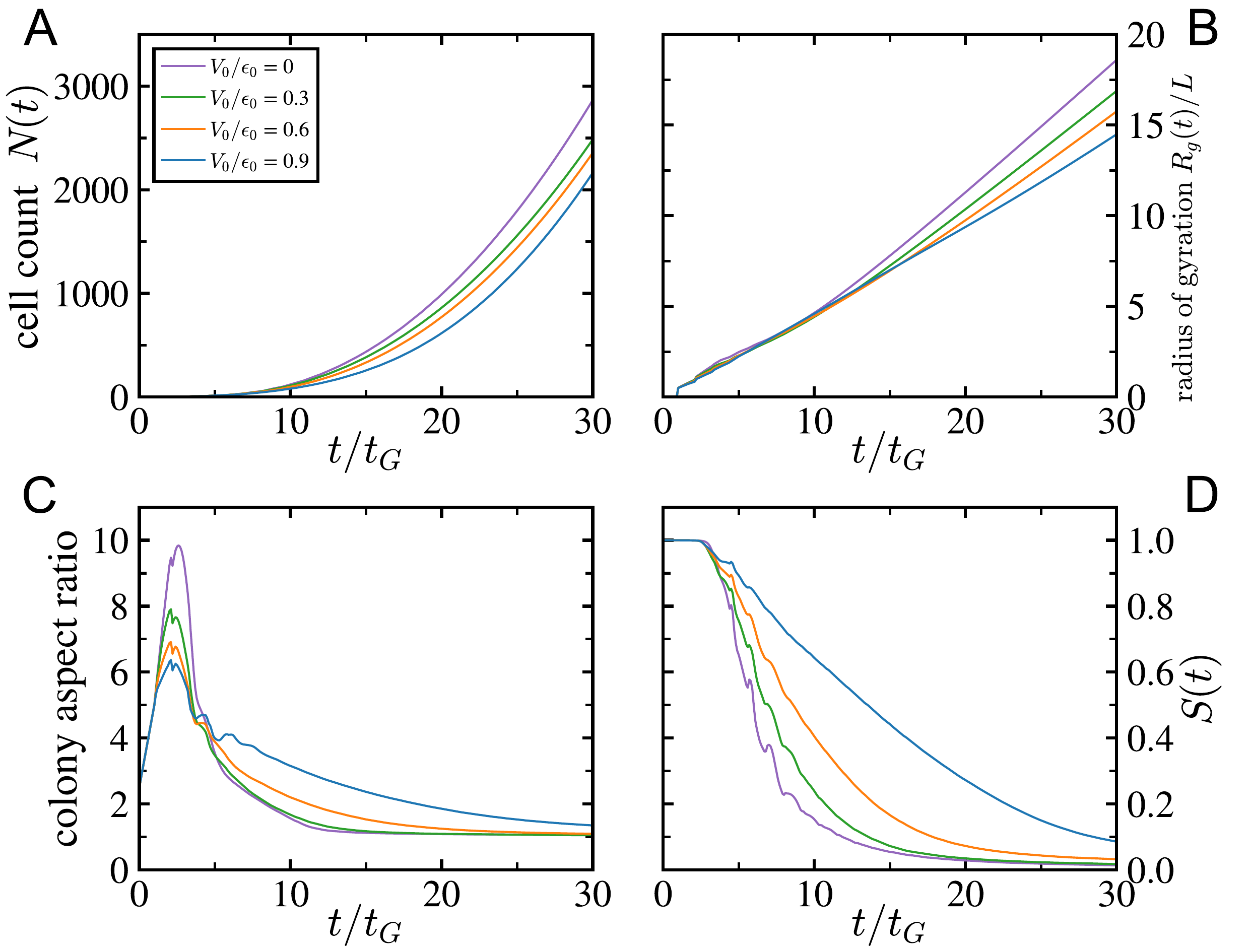}
    \caption{
    \textbf{Supplementary data of colony properties.} (A) Total number of particles $N(t)$ over time. (B) Radius of gyration $R_g(t)$ over time. (B) Colony aspect ratio calculated from the dimensions of the minimum bounding rectangle. (D) Global nematic order parameter $S(t)$ as a function of time.
    }
    \label{fig_SI}
\end{figure}

\subsection*{Supplementary data analysis}
The number of particles over time (Supplementary Fig.~\ref{fig_SI}(A)) confirms the behavior seen in the colony area.
The radius of gyration (Supplementary Fig.~\ref{fig_SI}(B)) provides another measure of colony size, showing a similar confinement effect for larger $V_0$.
We also calculated the aspect ratio using the colony's minimum bounding rectangle (Supplementary Fig.~\ref{fig_SI}(C)), a geometric measure that matches the result obtained by the gyration tensor but avoids the divergence at early times.

Finally, the global nematic orientational order parameter
\begin{equation}
    S(t) = |\langle \exp(2 \mathrm{i} \phi_j) \rangle|
\end{equation}
quantifies the overall alignment of the system at a given time $t$.
Here, $\phi_j$ represents the orientational angle of particle $j$ and the angled brackets $\langle \ldots \rangle$ denote the average over all particles $j$ present in the system at that specific time $t$.
This parameter ranges from $0$ (isotropic, no preferred orientation) to $1$ (perfect alignment).
The global nematic order parameter (Supplementary Fig.~\ref{fig_SI}(D)) shows a trend consistent with the average local nematic order (see main text).

To assess the local orientational order around each individual particle as provided in the main text, the local nematic orientational order parameter $S(\vec{r}_j,t)$ is calculated.
For each particle $j$ located at position $\vec{r}_j$ at time $t$,
\begin{equation}
    S(\vec{r}_j,t) = |\langle \exp(2 \mathrm{i} \phi_k) \rangle_{\vec{r}_j}|
\end{equation}
is determined by averaging over its surrounding neighbors.
In this formulation, the angled brackets $\langle \ldots \rangle_{\vec{r}_j}$ denote the average over all particles $k$ that are located within a local circular region centered at $\vec{r}_j$ with a specified radius of $3L$ \cite{monderkamp2021topology}.

The average local nematic order parameter $S_\text{avg}(t)$ provides a system-wide measure of the local order.
It is obtained by averaging the individual local nematic order parameters $S(\vec{r}_j,t)$ over all particles in the system at a given time, i.e.
\begin{equation}
    S_\text{avg}(t) = \frac{1}{N(t)}\sum_{j=1}^{N(t)} S(\vec{r}_j,t) \;.
\end{equation}

\begin{figure}
\centering
\includegraphics[width=\columnwidth] {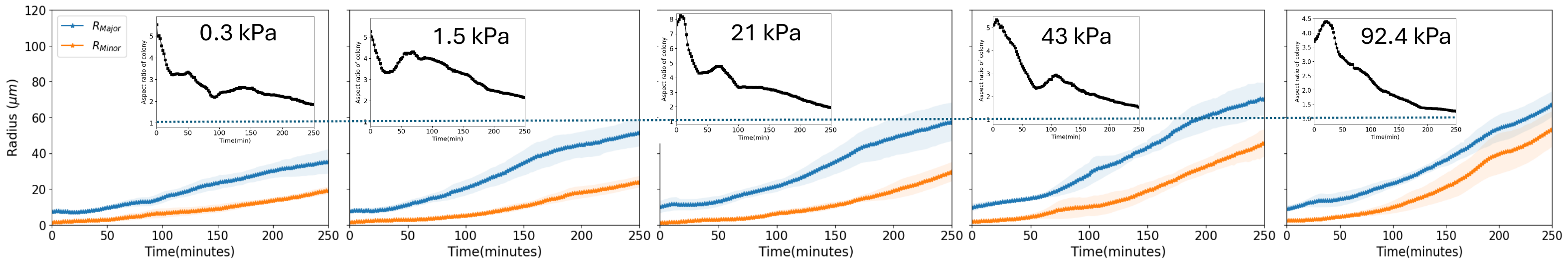}
\caption{\textbf{Colony-scale geometric features.} The major (blue) and minor (orange) axes are plotted as function of time, for the  substrates with increasing Young's modulus as mentioned in inset. Inset also shows the corresponding times series of colony aspect ratio in all cases.}
\label{fig:FIG2_MAIN}
\end{figure}

\begin{figure}[ht]
\centering
\includegraphics[width=0.6\columnwidth] {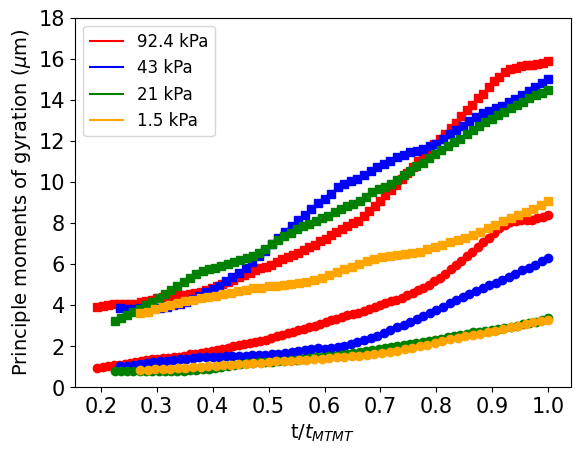}
\caption{\textbf{Principle moments of the gyration tensor derived from experiments.} Variation of the two eigenvalues of the gyration tensor over time. The minor eigen values are represented by the lower part of the plot while the major eigen values occupy the upper the part of the plot. In agreement with the simulations, variation of the minor eigen value is suppressed for softer substrates.}
\label{fig:GYRExp}
\end{figure}

\begin{figure}[ht]
\centering
\includegraphics[width=\columnwidth] {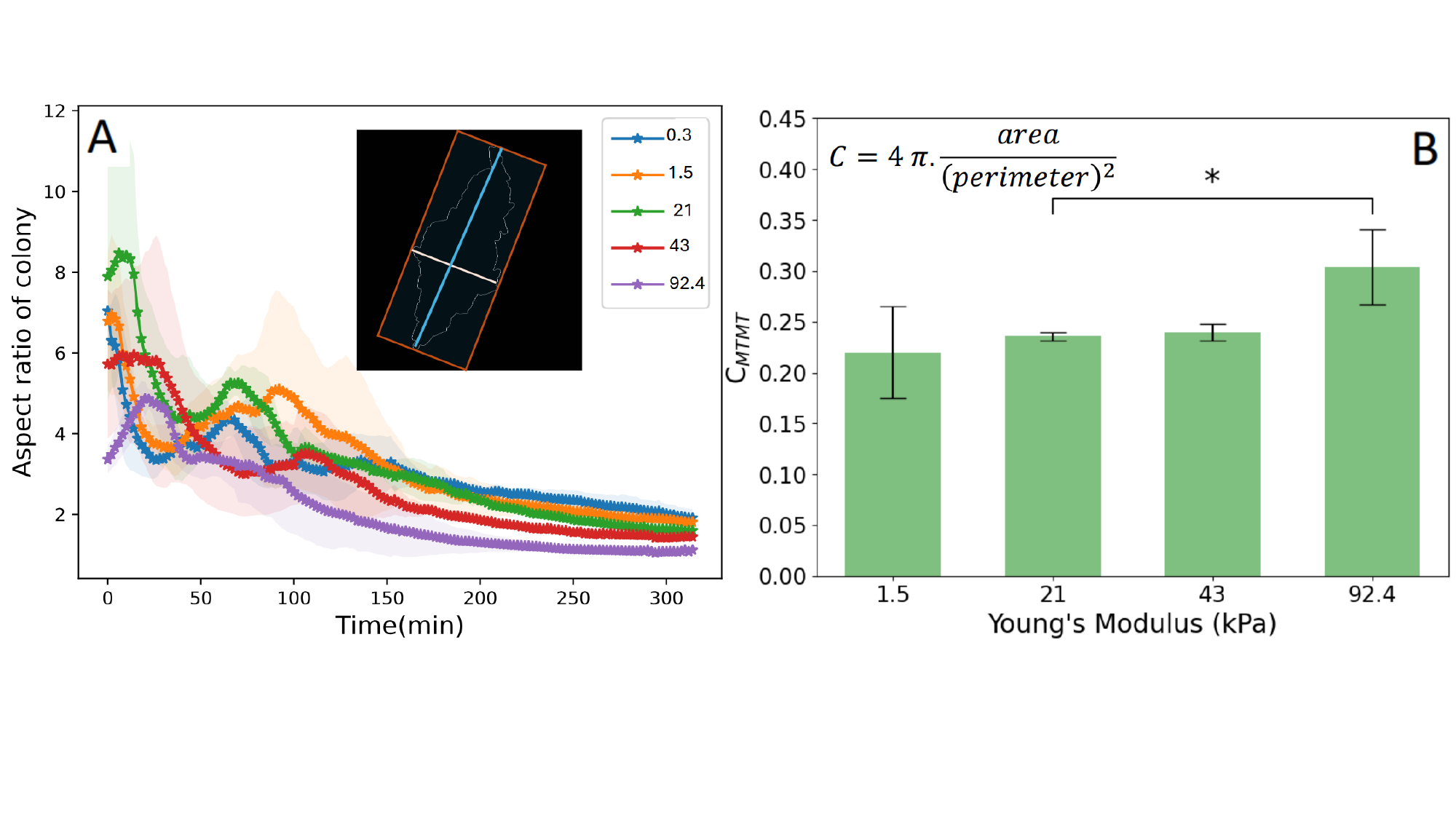}
\caption{\textbf{Shape anisotropy of the colonies.} (A) Colony aspect ratio is plotted as a function of time of colony growth for various substrates as labeled by their Young's modulus. Inset shows the bounding box on the colony contour to compute the major and minor distance of the colony. The mean (solid line) and standard deviation (shaded region) are calculated from at least three biological replicates for each stiffness condition. (B) Circularity (measures the compactness of the colony morphology) at transition time from single to multi-layered morphology, as substrate stiffness varies. A significant difference in mean colony radius, colony area and number of cells at the MTMT is seen when Young’s modulus varies from 1.5 kPa to high values 92.4 kPa (two sample t-test, p-value$<$ 0.01). Circularity values also display variation with the Young's modulus, between 21 kPa to 92.4 kPa (p-values $<$0.05).}
\label{fig:FIGS1}
\end{figure}

\begin{figure}[ht]
\centering
\includegraphics[width=4 in] {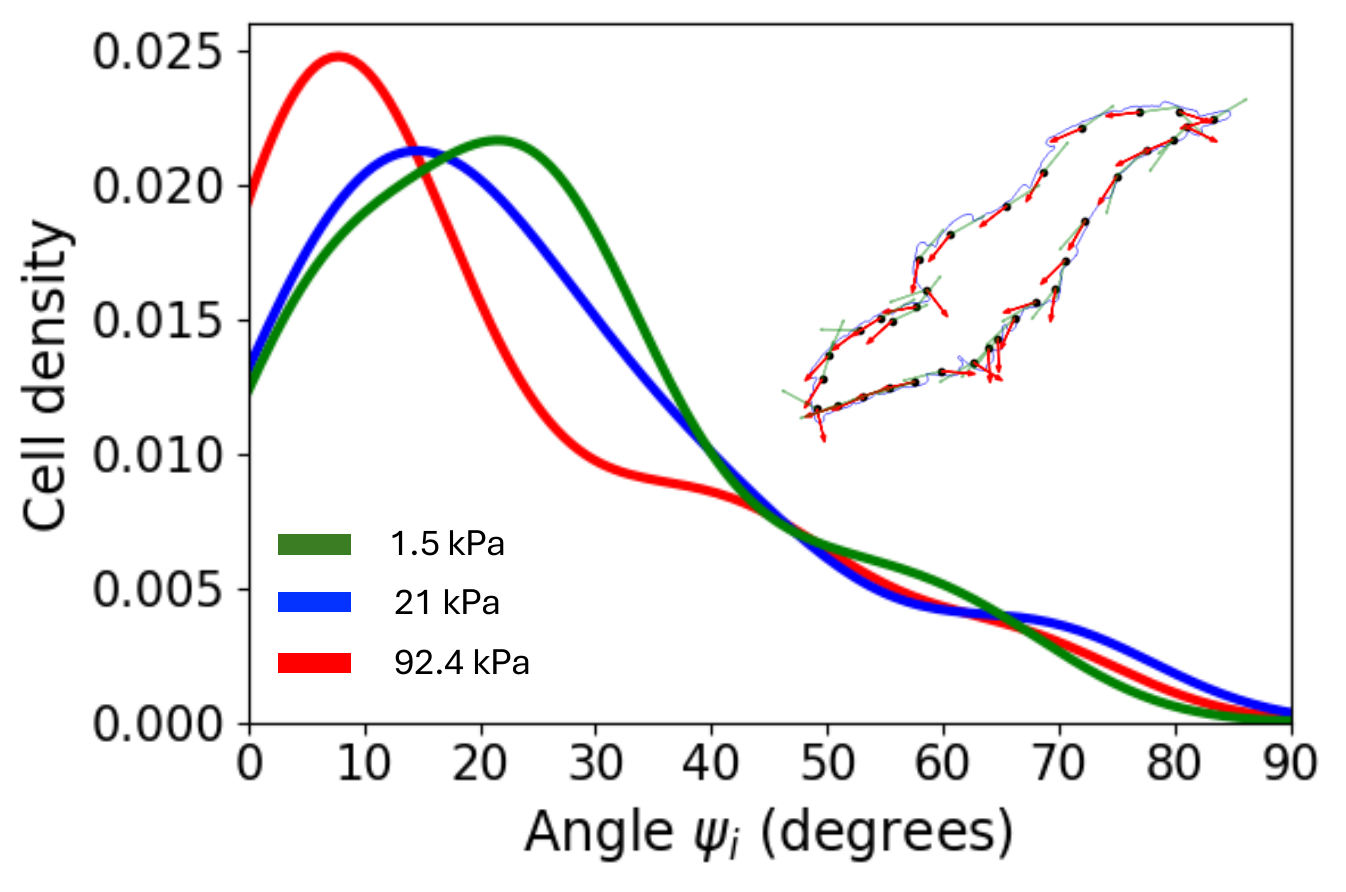}
\caption{\textbf{Cell alignment at the colony boundary.} The probability density of the acute angle between the boundary tangent (green arrows) and the orientation (red arrows) of bacterial cells (black centroids) reveals that most cells tend to align tangentially with the colony boundary across all conditions. However, substrate stiffness influences the extent of this alignment: cells growing on stiffer substrates exhibit a higher degree of tangential alignment compared to those on softer substrates.} 
\label{fig:FIGS31}
\end{figure}

\begin{figure}[ht]
\centering
\includegraphics[width=\columnwidth] {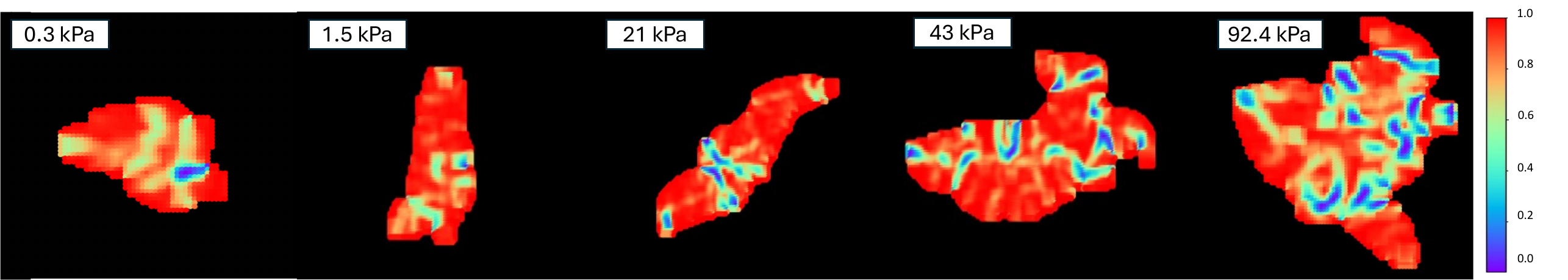}
\caption{\textbf{Nematic order parameter for colonies growing on substrates with different Young's modulii.} Color map of distribution of values of nematic order parameter $S_\text{avg}$ in shown for a representative case for colony at MTMT time growing on substrates with increasing Young's modulus from left to right (MTMT time of bacteria colonies: $t_c \approx 130$ min for 0.3 kPa;  $t_c \approx 160$ min for 1.5 kPa; $t_c \approx$ 200 min for 21 kPa and $t_c \approx 235$ min for 92.4 kPa substrate.) } 
\label{fig:FIGSOP}
\end{figure}

\begin{figure}[ht]
\centering
\includegraphics[width=4 in] {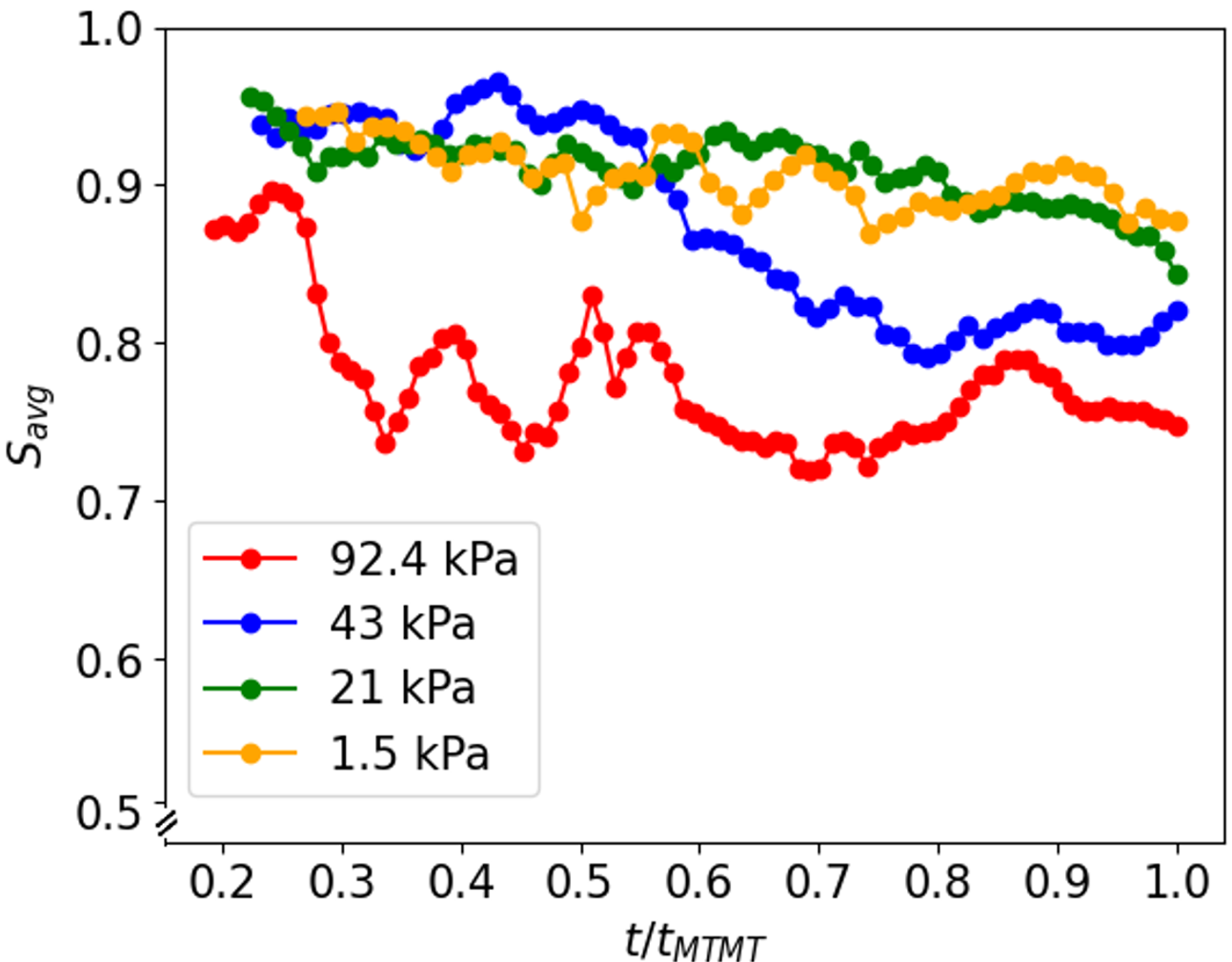}
\caption{\textbf{Time evolution of the average nematic order parameter.} Average nematic order parameter of colonies, $S_\text{avg}(t)$, growing on substrates with different Young's modulii. The overall orientational alignment is better preserved on the softer surfaces, in agreement with the agent-based simulation results.} 
\label{fig:OPTime}
\end{figure}

\begin{figure}[ht]
    \centering
    \includegraphics[width=1.0\linewidth]{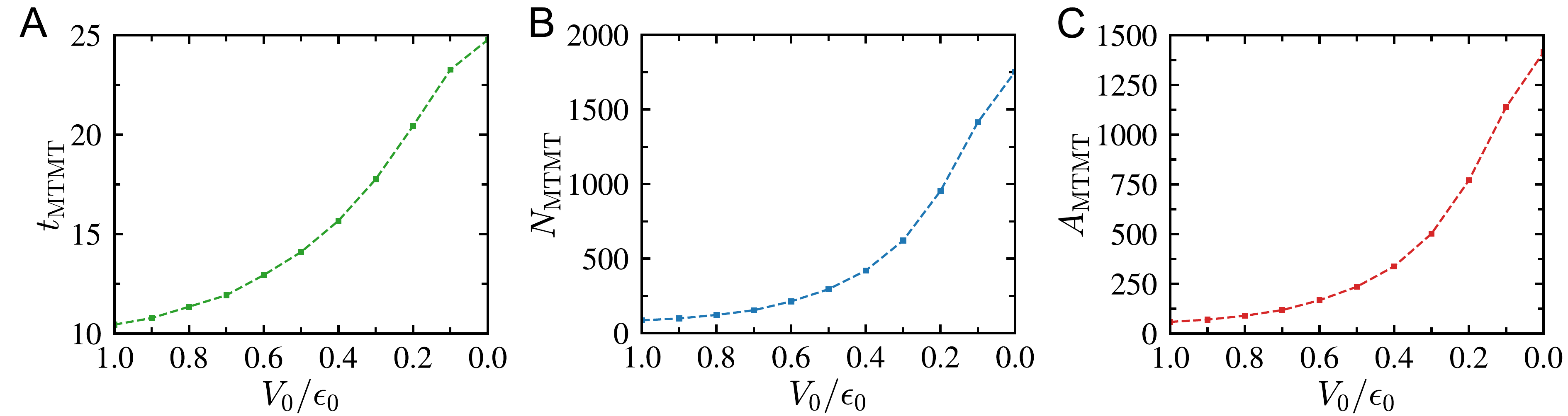}
    \caption{
    \textbf{Estimating the onset of the mono-to-multilayer transition.} (A) The time at which the multilayer transition is predicted to occur as a function of the softness parameter $V_0$. (B) The total number of particles in the colony at the transition point. (C) The total area of the colony at the transition point.
    }
    \label{fig_SI-2}
\end{figure}

\begin{figure}[ht]
\centering
\includegraphics[width=4in] {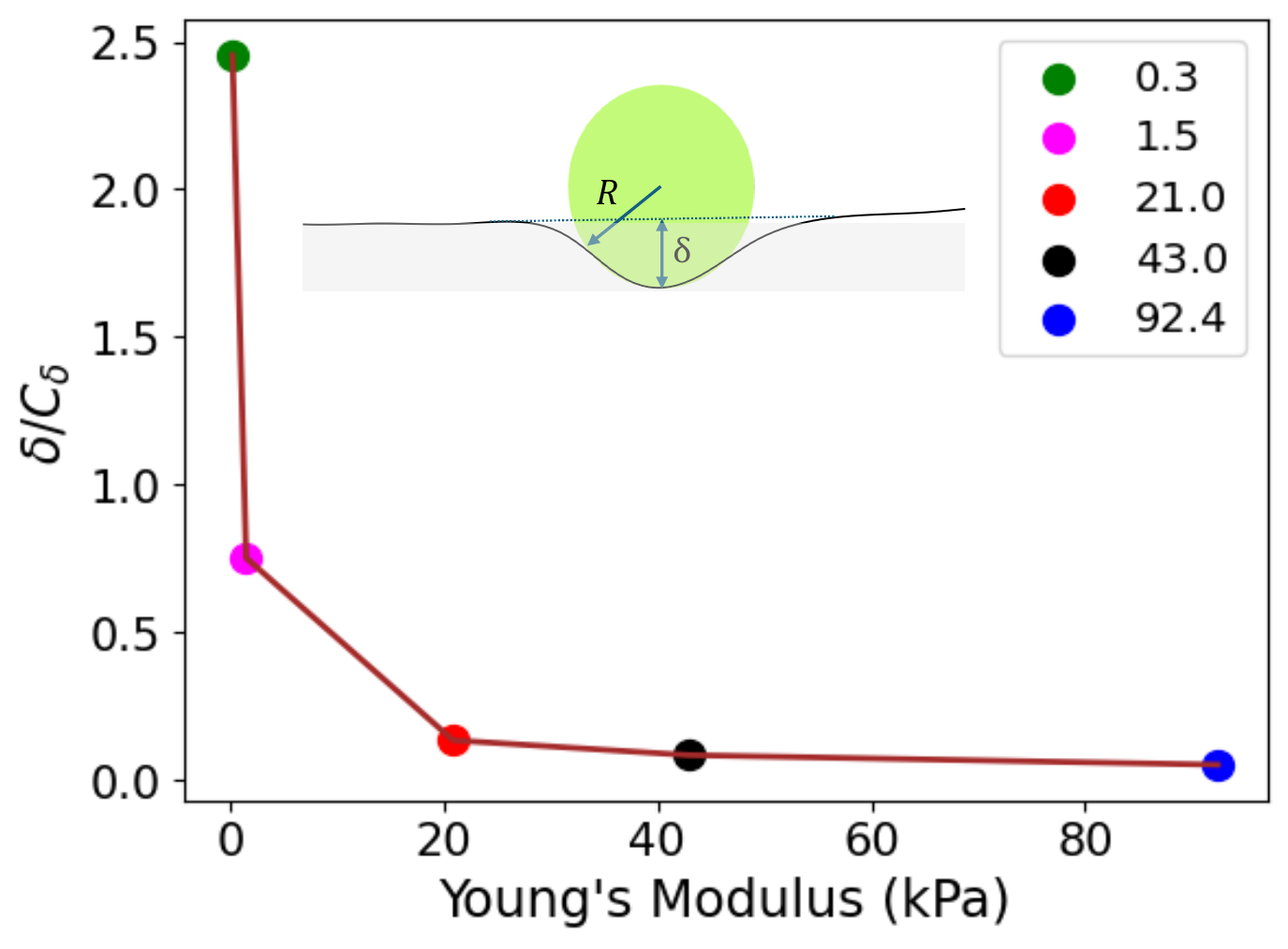}
\caption{\textbf{Emergence of effective drag forces on growing colonies.} Relative sinking depth estimated for colonies growing on the various substrates, with dots indicating the Young's modulus values. (Inset) Illustrative sinking of colony into substrate, with depth estimated using Hertzian contact.}
\label{fig:FIG7}
\end{figure}

\begin{figure}[ht]
\centering
\includegraphics[width=\columnwidth] {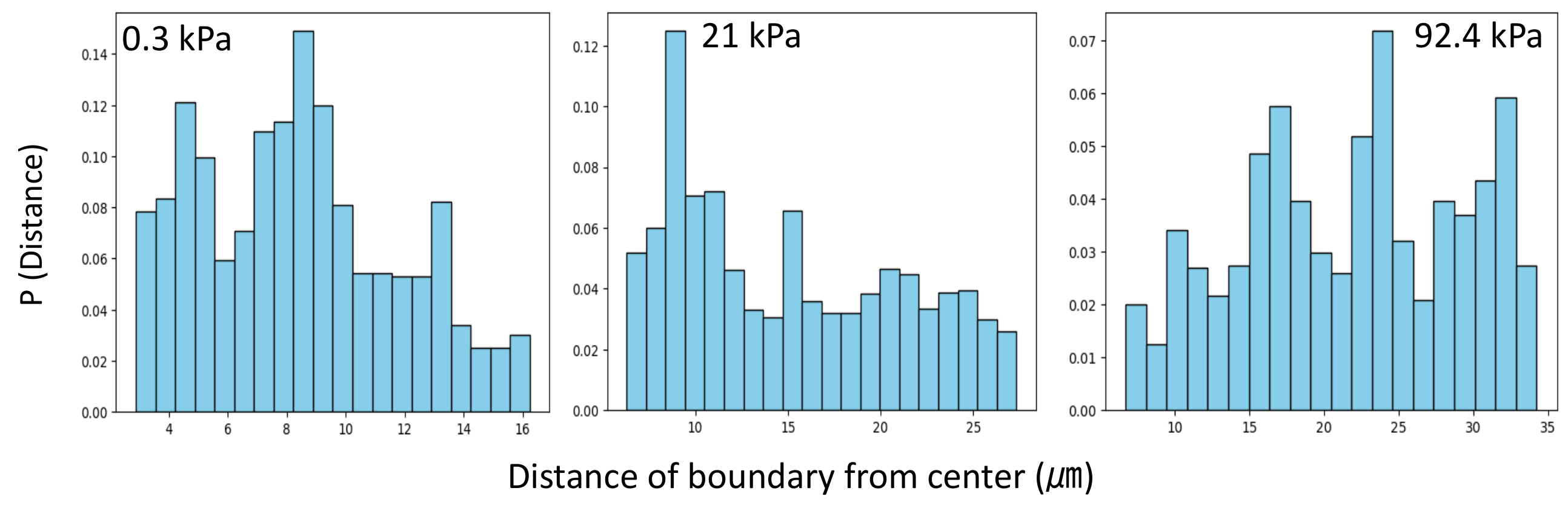}
\caption{\textbf{Boundary distance from the colony center.} Probability distribution of distance from the colony center across a $360^{\circ}$ angle for substrate of 0.3 kPa, 21 kPa and 92 kPa Young's modulus values, at MTMT. The mean value of distance from center ($\sim R_{mean}$) for different cases were $10 ~\mu m$, $17.5 ~\mu m$ and $22.5 ~\mu m$ respectively.}
\label{fig:FIGS3}
\end{figure}

% Put this into SI if desired

\bibliography{paper.bib}

\bibliographystyle{unsrt}